\documentclass{JHEP3}
\usepackage{booktabs}
\usepackage{graphicx}

\def\ccw{{\hspace{-2.5mm}\unitlength 0.1in
\begin{picture}(1.00,1.00)(8.70,-11.50)
\special{pn 8}%
\special{pa 1120 1120}%
\special{pa 1100 1100}%
\special{pa 1080 1120}%
\special{fp}%
\end{picture}%
\hspace{-0mm}}}

\def\cw{{\hspace{-2.5mm}\unitlength 0.1in
\begin{picture}(1.00,1.00)(8.70,-11.50)
\special{pn 8}%
\special{pa 1120 1100}%
\special{pa 1100 1120}%
\special{pa 1080 1100}%
\special{fp}%
\end{picture}%
\hspace{-0mm}}}

\def\bx{{\hspace{0.6mm}\unitlength 0.1in
\begin{picture}( 1.00,1.00)(10.00,-13.00)
\special{pn 8}%
\special{pa 1000 1210}%
\special{pa 1050 1210}%
\special{pa 1050 1260}%
\special{pa 1000 1260}%
\special{pa 1000 1210}%
\special{fp}%
\end{picture}%
\hspace{-0mm}}}

\def\bxx{{\hspace{0.6mm}\unitlength 0.1in
\begin{picture}(1.00,1.00)(10.00,-13.00)
\special{pn 8}%
\special{pa 1000 1210}%
\special{pa 1050 1210}%
\special{pa 1050 1260}%
\special{pa 1000 1260}%
\special{pa 1000 1210}%
\special{pa 1000 1260}%
\special{pa 1050 1260}%
\special{pa 1050 1310}%
\special{pa 1000 1310}%
\special{pa 1000 1260}%
\special{fp}%
\end{picture}%
\hspace{0mm}}}

\def\bxxx{{\hspace{0.6mm}\unitlength 0.1in
\begin{picture}(1.00,1.00)(10.00,-13.00)
\special{pn 8}%
\special{pa 1000 1210}%
\special{pa 1050 1210}%
\special{pa 1050 1260}%
\special{pa 1000 1260}%
\special{pa 1000 1210}%

\special{pa 1000 1260}%
\special{pa 1050 1260}%
\special{pa 1050 1310}%
\special{pa 1000 1310}%
\special{pa 1000 1260}%

\special{pa 1000 1310}%
\special{pa 1050 1310}%
\special{pa 1050 1360}%
\special{pa 1000 1360}%
\special{pa 1000 1310}%
\special{fp}%
\end{picture}%
\hspace{0mm}}}

\renewcommand{\sp}{p\hspace{-.42em}/}
\newcommand{\sq}{q\hspace{-.50em}/}
\newcommand{\E}{{\cal E}}

\newcommand{\beq}{\begin{equation}}
\newcommand{\eeq}{\end{equation}}
\newcommand\beqa{\begin{eqnarray}}
\newcommand\eeqa{\end{eqnarray}}
\newcommand\bea{\begin{array}}
\newcommand\eea{\end{array}}
\newcommand{\nn}{\nonumber}
\newcommand{\neqa}{\nonumber\end{eqnarray}}
\newcommand{\la}{\label}
\newcommand{\J}{{\cal J}}
\newcommand{\iint}{{\cal I}}

\renewcommand{\O}{{\cal O}}
\newcommand{\color}[1]{}

\newcommand{\ii}{{\rm i}}
\newcommand{\jj}{{\rm j}}

\newcommand{\sint}[2]{\int\limits_{#1}\hspace{-#2}-\hspace{#2}}

\newcommand{\eq}[1]{(\ref{#1})}
\newcommand{\eqs}[2]{(\ref{#1},\ref{#2})}

\renewcommand{\t}{\tilde}

\def\({\left(}
\def\){\right)}
\def\[{\left[}
\def\]{\right]}

\def\<{\langle}
\def\>{\rangle}

\def\d{\partial}

\def\sG{/\hspace{-0.25cm}G}
\def\sF{/\hspace{-0.25cm}F}
\def\sH{/\hspace{-0.25cm}H}
\def\scot{\cot\hspace{-.43cm}/\hspace{.14cm}}

\title{Complete 1-loop test of AdS/CFT}
\author{Nikolay Gromov\\Laboratoire de Physique Th\'eorique
de l'Ecole Normale Sup\'erieure et l'Universit\'e Paris-VI, Paris,
75231, France; St.Petersburg INP, Gatchina, 188 300, St.Petersburg,
Russia; E-mail: \email{nikgromov@gmail.com}}

\author{Pedro Vieira\\Laboratoire de Physique Th\'eorique
de l'Ecole Normale Sup\'erieure et l'Universit\'e Paris-VI, Paris,
75231, France;  Departamento de F\'\i sica e Centro de F\'\i sica do
Porto Faculdade de Ci\^encias da Universidade do Porto Rua do Campo
Alegre, 687, \,4169-007 Porto, Portugal; E-mail:
\email{pedrogvieira@gmail.com}}

\abstract{We analyze nested Bethe ansatz (NBA) and the corresponding finite
size corrections. We find an integral equation which describes these
corrections in a closed form. As an application we considered the
conjectured Beisert-Staudacher (BS) equations with the Hernandez-Lopez
dressing factor where the finite size corrections should reproduce
generic one (worldsheet) loop computations around any classical
superstring motion in the $AdS_5\times S^5$ background with exponential precision in the large angular momentum of the string states. Indeed, we show
that our integral equation can be interpreted as a sum over all
physical fluctuations and thus prove the complete $1$-loop consistency
of the BS equations. In other words we demonstrate that any local
conserved charge (including the AdS Energy) computed from the BS
equations is indeed given at 1-loop by the sum of charges of
fluctuations up to exponentially suppressed contributions. Contrary to all previous studies of finite size
corrections, which were limited to simple configurations inside rank
$1$ subsectors, our treatment is completely general. }

\preprint{arXiv:yymm.nnnn\\ LPTENS 07/47}

\begin{document}

\section{Introduction}


Bethe equations \cite{Bethe} describe the scattering of the fundamental
degrees of freedom of integrable $1+1$ dimensional theories defined on
some large circle of length $\cal L$. The existence of a large amount
of conserved charges results in the factorizability property of the
scattering matrix. Namely the full $n$ particle $S$-matrix is
completely fixed by the $2$ particle scattering. Moreover in two
dimensions this $2$ to $2$ scattering process conserves not only the
total momentum but also the set of individual momenta.  Then, for a
large enough circle, the momenta of the several particles are quantized
through the wave function periodicity condition
\beq
1=e^{i p_k {\cal L}}\prod_{j\neq k}^{L} S(p_k,p_j) \la{verysimple}
\eeq
meaning that the (trivial) phase acquired by a particle with momentum
$p_k$ while going around the circle equals the free propagation plus
the scattering phases shifts (or time delay in coordinate space) due to
the passage through each of the other particles. In general Bethe
equations are only asymptotically exact as ${\cal L}\to\infty$
otherwise wrapping effects \cite{warp1,warp2} must be taken into
account.

Equation \eq{verysimple} is, however, describing particles with no
isotopic degrees of freedom, that is $S(p_k,p_j)$ is just a phase. In
general, when we have some nontrivial symmetry group, this is not the
case and, rather, we must solve the diagonalization problem
$$
|\psi\rangle=e^{ip_k{\cal L}}\prod_{j\neq k}^{L}\ \hat S\(p_k,p_j\)
|\psi\rangle  \nn
$$
where $\hat S(p_k,p_j)$ is now a matrix and $\left|\psi\right\rangle$
is the multi-particle wave function (for integrable theories the number
of particles is conserved). If the scattered particles transform under
some symmetry group we will obtain not just one equation like
(\ref{verysimple}) but rather a set of $n$ equations entangling the
scattering of particles with momenta $p_k$ and $p_j$ in space-time with
the scattering of spin waves in the isotopic space.

In this paper we will mainly consider the particular limit of low
energies when the wave length of the spin waves are large and particles
exhibit collective behavior which, in some important cases, can be
associated with the classical motion of collective fields. By studying
carefully this limit one can get important information about the
quantization of some classical field theories.

In terms of the Bethe ansatz equation this corresponds to a limit,
first considered in the condensed matter literature by Sutherland
\cite{Sutherland} in the study of the ferromagnetic limit of the
Heisenberg chain and rediscovered and generalized in the context of
AdS/CFT
\cite{umbrella}, where the Bethe roots $u_j\sim \cot(p_j/2)$ scale
with the number of such roots and with the total number of particles,
$u_j\sim K_a \sim L$. In this limit the Bethe roots condense into
disjoint cuts. Since there are several types of Bethe roots, one for
each Bethe equation, the condensation of the Bethe roots for systems
with $n$ Bethe equations will generate some Riemann surface with $n+1$
sheets as in figure \ref{Riemann}. This resulting curve is in
one-to-one correspondence with the curves classifying classical
solutions through the finite gap method
\cite{book1,book2,KMMZ,BKSZ,GKSV}. In this way one finds the
semi-classical spectrum of the theory.

The next natural step is to compute the first quantum corrections to
the semi-classical spectrum, which from the Bethe ansatz point of view
will correspond to the finite size (i.e. $1/L$) corrections. For the
simplest Bethe equations of the form (\ref{verysimple}) these
corrections, called anomalies, were known \cite{FS1,FS3,FS4,FS5,BF,GK,r17}
but for nested Bethe ansatz equations the analysis is much more
delicate due to the formation of bound states, called stacks
\cite{BKSZII}, which are the basic constituents of the cuts made out of
more than one type of Bethe roots like the ones in figure
\ref{fig:2conf}. In this paper we develop the necessary tools to deal
with these richer systems with isotopic degrees of freedom.

Particularly important tools are the so called
\textit{dualities}. One of them, the \textit{fermionic} duality, is
well studied
\cite{FD1,FD2,FD3,FD4,FD5,BKSZII,KSZ} and has a clear mathematical
meaning. If the symmetry group under which the fundamental particles
transform is a super group then there are several possible choices of
NBA equations corresponding to the several possible choices of super
Dynkin diagram which, for super-groups, is not unique. These equations
are related by some dualities associated with the fermionic nodes of
the corresponding super Dynkin diagram. In the scaling limit they
correspond to the exchange of Riemann sheets. In this paper we also use
an analogue of this duality, baptized \textit{bosonic} duality, which
exists even in the case of a purely bosonic symmetry. It is associated
with the bosonic nodes of the Dynkin diagram.


Next we apply our method to the recently conjectured Beisert-Staudacher
(BS) Bethe equations \cite{BS}. These equations contain a free
parameter $\lambda$ and should describe two systems at the same time:
four dimensional $\mathcal{N}=4$ SYM and type IIB super-strings in
$AdS_5\times S^5$, two theories which are conjectured to be dual
\cite{Maldacena:1997re,Gubser:1998bc,Witten:1998qj}. At weak coupling,
$\lambda \ll 1$, we are in the perturbative regime of $\mathcal{N}=4$
SYM and the Bethe equations describe the spectrum of the planar
dilatation operator which can be considered as a spin chain Hamiltonian
\cite{MZ,B} with $PSU(2,2|4)$ symmetry. At strong
coupling $\sqrt{\lambda} \sim L \gg 1$ the theory describes classical
super-strings in the curved space-time $AdS_5\times S^5$
\cite{KMMZ,BKSZII} and the $1/\sqrt{\lambda}$ corrections in the
scaling limit correspond to the semi-classical quantization of such
highly non-trivial quantum field theory.

As it was stressed in our previous papers~\cite{GV1,GV2} there are two
completely different ways to compute the 1-loop correction to the
quasi-classically quantized energy spectrum. One, straightforward but
technically more involved, is to take the NBA equations, compute its
spectrum and then expand it in powers of $1/\sqrt\lambda$ i.e. find its
finite size corrections. Another way, more indirect one, is to pick
some classical solution satisfying the semi-classical quantization
condition, and quantize around it, i.e. find the spectrum of all
possible excitations of this solution. The one loop shift is then given
by the zero energy oscillations and is equal to half of the graded sum
of all excitation energies, like for a simple set of independent one
dimensional harmonic oscillators.

Both calculations can be performed using the BS equations and it is a
very nontrivial test of the proposed equations that these two
calculations give the same result. In fact for the second calculation
we do not even need the Bethe ansatz, since it is based completely on
the semi-classical quantization which, as shown in
\cite{GV1}, can be performed relying uniquely on the classical integrable structure of the theory -- the algebraic curve \cite{KMMZ,BKSZ}. Moreover we expect the
second approach to give the exact result whereas the first one is only
valid as long as one can trust the asymptotic BAE, which suffers from
the wrapping effects mentioned above. Indeed we found that the two
results coincide not precisely, but only for large $L/\sqrt\lambda$
with exponential precision. This is obviously a manifestation of the
wrapping effects considered in the $AdS/CFT$ context in
\cite{J1,r34,warp3,warp4,J2}. This exponential mismatch was first observed in \cite{r38}.

\FIGURE[t]{
    \centering
        \resizebox{70mm}{!}{\includegraphics{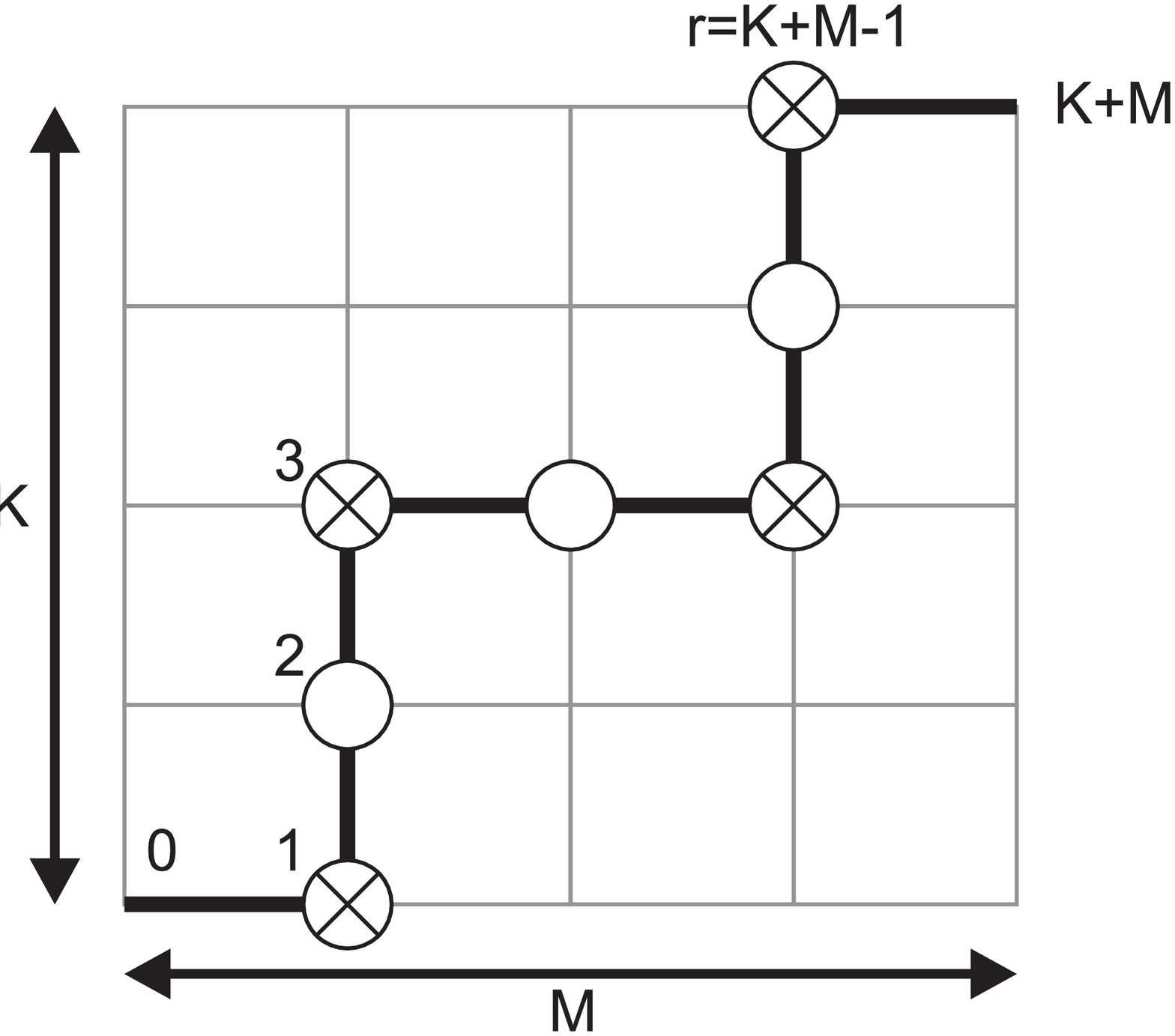}}
        \caption{For $su(K|M)$ super algebras the Dynkin diagram is not
        unique. The several possible choices can be represented as
        the paths going from the up right corner $(M,K)$ to the origin
        always approaching this point with each step. The turns are the fermionic
        nodes whereas the straight lines correspond to the usual bosonic
        nodes. Different paths will correspond to different sets of Bethe
        equations which are related by fermionic dualities which flip a
        \textit{left--down} fermionic turn into \textit{down--left} turn
         or vice-versa \cite{KSZ}.\la{figH0}}
}

Finally we should stress that we follow a constructive approach. That
is we start from the classical integrable structure, the finite gap
curves. The curves can be described by some integral equations. We find
how to correct this equations
\textit{in such a way} that they will now describe not only the
classical limit $\sqrt\lambda\to\infty$ but also the $1/\sqrt\lambda$
corrections. Then we show that the integral equations modified  in this
way coincide precisely with the scaling limit expansion of the BS
equations \cite{BS} with the Hernandez-Lopez phase \cite{HL} (up to some exponentially suppressed wrapping effects,
irrelevant for
 large angular momentum string states)!
Our comparison, being done at the functional level, is completely
general.

This paper is organized as follows. In section \ref{one} we introduce
some notations, the notion of stack and the bosonic duality. In section
\ref{anomalies} we derive, in two independent ways, an integral
equation describing the finite size corrections to the leading limit -
using the dualities and using the transfer matrices.  In section
\ref{BSsection} we follow the constructive approach mentioned above to
re-derive the same integral equation from the equations in the scaling
limit. Section \ref{dualitysec} contains some details about the bosonic
duality such as some theorems and examples -- the reader interested
only in the main results of the paper can skip this section. In section
\ref{BSsection} we apply the methods developed in the previous sections
to the study of the BS equations, compute the finite size corrections
and relate them with the quantum fluctuations of the theory. Appendix A
is devoted to the study of the invariance of the transfer matrices of
$su(K|M)$ supergroups under the bosonic duality and in Appendix B we
derive an integral equation describing the semi-classically corrected
equations for $su(n)$ spin chains.

\section{Nested Bethe Ansatz and Bosonic Duality} \la{one}
In the first sections we stick mainly to the simple example of
$su(1,2)$ spin chain, although our main motivation comes from its
application to $AdS_5\times S^5/\mathcal{N}=4$ SYM correspondence where
the symmetry group is $PSU(2,2|4)$. Indeed this simple toy model
contains already all the nontrivial new features appearing due to the
Nested nature of the Bethe ansatz.  The generalization to other
(super)groups is straightforward and in particular we shall focus on
the Bethe ansatz describing the superstring in $AdS_5\times S^5$ in
section
\ref{BSsection}.

For integrable rank $r$ spin chains each quantum state is parameterized
by a set $\{u_{a,j}\}$ of Bethe roots where $a=1,\dots,r$ refers to the
Dynkin node and $j=1,\dots,K_a$ where $K_a$ is the excitation number of
\textit{magnons} of type $a$. The Bethe equations from which we find
the position of these roots are then given by
\beq
e^{i\tau_a}\(\frac{u_{a,j}+\frac{i}{2} V_a}{u_{a,j}-\frac{i}{2}
V_a}\)^L=-\prod_{b=1}^{r}\frac{Q_b\(u_{a,j}+\frac{i}{2}
M_{ab}\)}{Q_b\(u_{a,j}-\frac{i}{2} M_{ab}\)} \la{BAEproduct}
\eeq
where
$$
Q_a(u)=\prod_{j=1}^{K_a}\(u-u_{a,j}\)
$$
are the Baxter polynomials, $V_a$ are the Dynkin labels of the
representation considered and $M_{ab}$ the Cartan matrix. In fact,
contrary to what happens for the usual Lie algebras, for super algebras
the Dynkin diagram (and thus the Cartan matrix) is not a unique. Take
for example the $su(K|M)$ super algebra. The different possible Dynkin
diagrams can be identified \cite{KSZ} as the different paths starting
from $(M,K)$ and finishing at $(0,0)$ (always approaching this point with each step) in a rectangular lattice of size $M\times K$ as in
figure \ref{figH0}. The turns in this path represent the fermionic
nodes whereas the bosonic nodes are those which are crossed by a
straight line -- see figure \ref{figH0} (the index $a$ goes along the
path as indicated). The Cartan matrix $M_{ab}$ is then given by
$$
M_{ab}=\(p_a+p_{a+1}\) \delta_{ab}-p_{a+1} \delta_{a+1,b}-p_{a}
\delta_{a,b+1}
$$
where $p_a$ is associated with the link between the node $a$ and $a+1$
and is equal to $+1$ ($-1$) if this link is vertical (horizontal).

Here we are considering twisted (quasi-periodic) boundary conditions
which, from an algebraic Bethe ansatz point of view corresponds to the
diagonalization of a transfer matrix with the insertion, inside the
trace, of an additional diagonal matrix \cite{Zab} which can be
parameterized by
\beq
g={\rm
diag}\,\(e^{i\phi_1},\dots,e^{i\phi_K},e^{i\varphi_1},\dots,e^{i\varphi_M}\)
\in SU(K|M) \la{g}
\eeq
and the twists $\tau_a$, appearing in (\ref{BAEproduct}) and associated
to a Dynkin node located at $(m,k)$ in the $M\times K$ network depicted
in figure \ref{figH0}, are then given by \cite{Zab}
\begin{eqnarray*}
\begin{array}{ll}
\tau_a=\phi_k-\phi_{k+1} & \textit{for a bosonic along a vertical segment of the path}\\
\tau_a=\varphi_{m+1}-\varphi_{m} & \textit{for a bosonic along a horizontal segment
of the path}\\
\tau_a=\varphi_{m+1}-\phi_{k}+\pi & \textit{for a fermionic node in a $\Gamma$ like
 turn that is with $p_{a-1}=-p_a=1$}\\
\tau_a=\phi_{k+1}-\varphi_{m}+\pi  & \textit{for a fermionic node with $p_{a-1}=-p_a=-1$}
\end{array}
\end{eqnarray*}
Notice that since $g \in SU(K|M)$ we have $\sum_k \phi_k-\sum_m
\varphi_m=0$ mod $2\pi$. We shall study these Bethe equations with
generic twists and we will see that the usual case ($\tau_a=0$) is in
fact quite degenerate.

As mentioned above, we find already all the ingredients we will need
for the study of the
 BS equations in the simple example of a $su(1,2)$ spin chain in the fundamental
  representation described by the following system of NBA equations\footnote{These
   equations are exactly the same as for the $su(3)$
spin chain except for the sign of the Dynkin labels which makes the
system simpler because the Bethe roots are in general real.}
\beqa
e^{i\phi_1-i\phi_2}&=&- \frac{Q_1\(u_{1,j}+i\)}{Q_1\(u_{1,j}-i\)}\frac{Q_2\(u_{1,j}-i/2\)}{Q_2\(u_{1,j}+i/2\)} \,\,\, , \,\,\, j=1\dots K_1 \la{eq1} \\
e^{i\phi_2-i\phi_3}\(\frac{u_{2,j}-\frac{i}{2} }{u_{2,j}+\frac{i}{2}
}\)^L&=&-\frac{Q_2\(u_{2,j}+i \)}{Q_2\(u_{2,j}-i
\)}\frac{Q_1\(u_{2,j}-i/2\)}{Q_1\(u_{2,j}+i/2\)}  \,\,\, , \,\,\, j=1\dots
K_2 \la{eq2}
\eeqa
The eigenvalues of the local conserved charges are functions of the roots $u_{2,j}$ only and are given by
\beq
{\cal
Q}_r=\sum_{j=1}^{K_a}\frac{i}{r-1}\(\frac{1}{(u_{2,j}+i/2)^{r-1}}-\frac{1}{(u_{2,j}-i/2)^{r-1}}\)\;.\la{su12charges}
\eeq
We will often denote these roots carrying charges by \textit{middle
node} roots.

First, consider only middle node excitations, $K_1=0\neq K_2$ in the
Sutherland scaling limit
\cite{Sutherland} where $u \sim K_2 \sim L\gg 1$. We shall always use
$x_{a,j}=u_{a,j}/L$ to denote the rescaled Bethe roots in the scaling
limit. Then, the Bethe equations in log form, to the leading order, can
be cast as
\beq
2\pi n_{j}+\phi_2-\phi_3=\frac{1}{x_{2,j}}+\frac{2}{L}\sum_{k\neq
j}\frac{1}{x_{2,j}-x_{2,k}} \la{discBAE}
\eeq
where the integers $n_j$ come from the choice of the branch of the
logs.
\FIGURE[t]{
    \centering
        \resizebox{151mm}{!}{\includegraphics{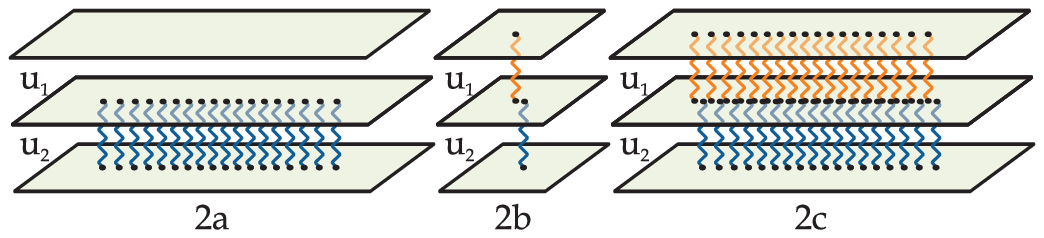}}
   \caption{
   The \textit{middle node} Bethe roots $u_2$ can condense
   into a line as depicted in figure \ref{fig:2confR}a
   (The spins in this spin chain transform in a non-compact
   representation and thus the cuts are topically real.
   For the $su(2)$ Heisenberg magnet the solutions are distributed in the complex plane as some \textit{umbrella} shaped curves \cite{umbrella}.). Roots of
   different types can form bound states, called stacks \cite{BKSZII}, as shown
   in figure \ref{fig:2confR}b. The stacks behave as fundamental excitations and
    can also form cuts of stacks as represented in figure \ref{fig:2confR}c.\label{fig:2confR}
   }
}

We see that we can think of the Bethe roots as positions of 2d Coulomb
charges on a plane with an external potential equal for every particle
plus an external force $2\pi n_j$ specific of each Bethe root. Thus, if
we group the $K_2$ mode numbers $\{n_{j}\}$ into $N$ large groups of
identical integers and consider the limit where both $L$ and $K_2$ are
very large with $K_2/L$ fixed, the Bethe roots will be distributed along
$N$ (real) cuts $\mathcal{C}^A$, each parameterized by a specific mode
number $\{n^A\}$ where $A=1,\dots,N$. Then the equations
(\ref{discBAE}) can be written through the density of \textit{middle
node} roots $x_2$ as
\beq
2\pi n^{A}+\phi_2-\phi_3=\frac{1}{x}+2\,\,\sG_2(x) \,\, , \,\, x \in
\mathcal{C}^A \la{contBAE}
\eeq
where we introduce the resolvents
\beq
\la{defdens}G_a(x)=
\int\limits\frac{\rho_a(y)}{x-y} \,\,\,\, , \,\,\,\, \rho_{a}(y)=\frac{1}{L}\sum_{j=1}^{K_a} \delta(x-x_{a,j})
\eeq
and where the slash of some function means the average of the function
above and below the cut,
$\sG(x)=\frac{1}{2}\(G(x+i\epsilon)+G(x-i\epsilon)\)$. Let us also
introduce some notation useful for what will follow. Defining the
quasi-momenta as
\beqa
p_1&=&-\frac{1}{2x}+G_1\hspace{9.4mm}-\phi_1 \,,\nn\\
p_2&=&-\frac{1}{2x}-G_1+G_2 -\phi_2\,, \la{ps}\\
p_3&=&-\frac{3}{2x}\hspace{9.4mm}-G_2-\phi_3 \,,\nn
\eeqa
we can add the indices $23$ to the mode number $n^A$ and to the cut
$\mathcal{C}^A$ in (\ref{contBAE}) and recast this equation as
\beq
2\pi n_{23}^{A}=\sp_2-\sp_3 \,\, , \,\, x \in \mathcal{C}_{23}^A   \,.
\la{P23}
\eeq
\FIGURE[t]{
    \centering
        \resizebox{151mm}{!}{\includegraphics{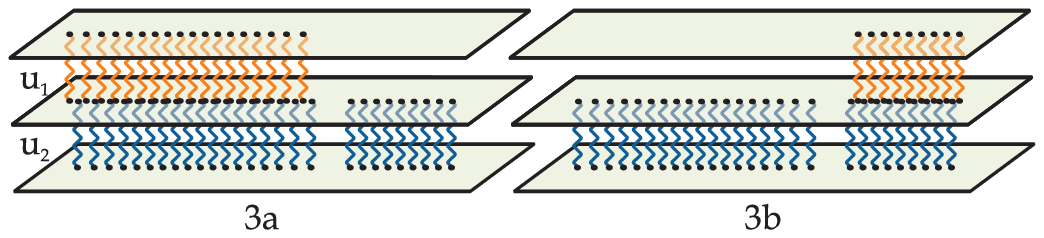}}
    \caption{In the scaling limit, to the leading order, the bosonic duality
    reads $Q_2\simeq Q_1 \t Q_1$ with $Q_a=\prod_{k=1}^{K_a}(u-u_a)$. Thus, if
     we start with the configuration in figure \ref{fig:2conf}a where the $K_1$
      roots $u_1$ form a cut of stacks together with $K_1$ out of the $K_2$ middle
       node roots $u_2$ and apply the bosonic duality to this configuration, the
        $K_2-K_1$ new roots $\t u_1$ must be close to the roots $u_2$ which were
         previously \textit{single} while the cut of stacks in the left of figure
          \ref{fig:2conf}a will become, after the duality, a cut of simple roots
           -- see figure \ref{fig:2conf}b.}\label{fig:2conf}
}

Next let us consider a state with only two roots $u_{2,1} \equiv u$ and
$u_{1,1} \equiv v$ with different flavors, that is $K_1=K_2=1$. Bethe
equations then yield
\beq
u=\frac{1}{2}\cot\frac{\phi_1-\phi_3+2\pi n}{2L} \,\, ,
\,\,v=u+\frac{1}{2}\cot\frac{\phi_1-\phi_2 }{2} \la{cot/2}
\eeq
which tells us that if $n\sim 1$ we are in the scaling limit where
$v\sim u\sim L$ and $v=u+\mathcal{O}(1)$ -- the two Bethe roots form a
bound state, called stack \cite{BKSZII}, and can be thought of as a
fundamental excitation -- see figure $2b$. On the other hand we notice
that, strictly speaking, for the usual untwisted Bethe ansatz with
$\phi_a=0$ the stack no longer exists.

Since the stack in figure $2b$ seems to behave as a fundamental
excitation one might wonder whether there exists a cut with $K_1=K_2$
roots of type $u_1$ and $u_2$, like in figure $2c$, \textit{dual} to
the configuration plotted in figure $2a$. To answer affirmatively to
this question let us introduce a novel kind of duality in Bethe ansatz
which we shall call \textit{bosonic duality}.

Indeed, as we explain in detail in section \ref{dualitysec}, given a
configuration of $K_1$ roots of type $u_1$ and $K_2$ roots of type
$u_2$, we can write
\beq
2i\sin\(\tau/2\) Q_2(u)=e^{i\tau/2}Q_1(u-i/2)\t Q_1(u+i/2)-e^{-
\tau/2}Q_1(u+i/2)\t Q_1(u-i/2)\,,
\la{duality}
\eeq
where
$$
\t Q_1(u)=\prod_{j=1}^{\t K_1}\(u-\t u_{1,j}\) \,\, , \,\, \t K_1=K_2-K_1\,,
$$
and $\tau=\phi_1-\phi_2$. Moreover this decomposition is unique and
thus defines unambiguously the position of the new set of roots $\tilde
u_1$. Then, as we explain in section \ref{dualitysec}, the new set of
roots $\{\t u_1,u_2\}$ is a solution of the same set of Bethe equations
(\ref{BAEproduct}) with
$$
\phi_1 \leftrightarrow \phi_2  \,.
$$
Let us then apply this duality to a configuration like the one in
figure $2a$ where the roots $u_2\sim L$ are in the scaling limit and
where there are no roots of type $u_1$, $K_1=0$. To the leading order,
we see that the $\tilde u_1$ in (\ref{duality}) will scale like $L$ so
that the $\pm i/2$ inside the Baxter polynomials can be dropped and we
find $Q_2\simeq
\tilde Q_1$, that is
$$
\tilde u_{1,j} = u_{2,j}+\mathcal{O}(1)
$$
and therefore we will indeed obtain a configuration like the one
depicted in figure $2c$. Moreover the local charges \eq{su12charges} of
this dual cut are exactly the same as those of the original cut $2a$
since they are carried by the
\textit{middle node} roots $u_2$
which are untouched during the duality transformation.

\FIGURE[t]{
    \centering
        \resizebox{151mm}{!}{\includegraphics{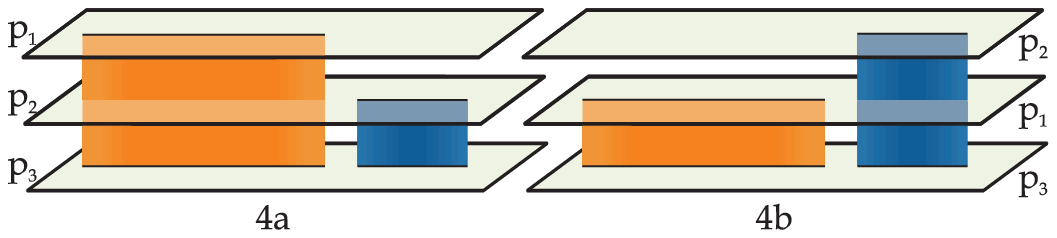}}
    \caption{
    In the scaling limit the configurations in figure \ref{fig:2conf} condense
    into some disjoint segments, cuts, and we obtain a Riemann surface whose
    sheets are the quasi-momenta. In this continuous limit the duality
    corresponds to the exchange of the Riemann sheets.
    }\label{Riemann}
}

Finally, if we apply the duality transformation to some configuration
like that in figure $3a$ in the scaling limit we find, by the same
reasons as above, that $Q_2(u)\simeq Q_1(u)\tilde Q_1(u)$.  This means
that the dual roots $\t u_1$ will be close to the roots $u_2$ which are
not yet part of a stack -- the ones making the cut in the right in
figure $3a$. Thus, after the duality, we will obtain a configuration
like the one in figure $3b$.

We conclude that, in the scaling limit with a large number of roots,
the distributions of Bethe roots condense into cuts in such a way that
the quasi-momenta $p_\ii$ introduced above become the three sheets of a
Riemann surface, see figure $4a$, obeying
\beq
2\pi n_{\ii\jj}^{A}=\sp_\ii-\sp_\jj \,\, , \,\, x \in
\mathcal{C}_{\ii\jj}^A \la{scaling} \,.
\eeq
when $x$ belongs to a cut joining sheets $i$ and $j$ with mode number
$n_{\ii\jj}^A$. The duality transformation amount to a reshuffling of
sheets $1$ and $2$ of this Riemann surface\footnote{As we shall see in
the next section this interpretation can be made exact, and not only
valid in the scaling limit.} so that a surface like the one plotted in
figure $4a$ transforms into the one indicated in figure $4b$.

\section{Anomalies -- finite size correction to Nested Bethe Ansatz equations} \la{anomalies}
In this section we will study the leading $1/L$ corrections to the
scaling equations (\ref{scaling}). Moreover since the charges of the
solutions are expressed through \textit{middle node roots} $u_2$ and
since these roots are duality invariant it is useful to write the Bethe
equations in terms of these roots only. Let us then consider a given
configuration of roots condensed into some simple cuts
$\mathcal{C}_{23}$ and some cuts of stacks $\mathcal{C}_{13}$. Then, to
leading order, at cuts $\mathcal{C}_{23}$ we have
\beqa
\frac{1}{x}+2 \int\limits_{\mathcal{C}_{23}}\hspace{-4.5mm}-\hspace{3mm} \frac{\rho_2(y) dy}{x-y} +\int\limits_{\mathcal{C}_{13}} \frac{\rho_2(y) dy}{x-y}=2 \pi n_{23}^A +\phi_2-\phi_3\,\, , \,\, x\in  \mathcal{C}_{23} \la{BAEmn}
\eeqa
because in a cut $\mathcal{C}_{13}$ we have $\rho_1\simeq
\rho_2+\mathcal{O}\(1/L\)$.
 To study finite size corrections to this equation two contributions must be
  considered. On the one hand when expanding the \textit{self interaction}
  we get \cite{FS1,FS3,FS4,FS5,BF,GK}
$$
\sum_{j\neq k}i \log\frac{u_{2,k}-u_{2,j}-i}{u_{2,k}-u_{2,j}+i}=2 \int\limits_{\mathcal{C}_{23}}\hspace{-4.5mm}-\hspace{3mm} \frac{\rho_2(y) dy}{x-y}+2\int\limits_{\mathcal{C}_{13}} \frac{\rho_2(y) dy}{x-y}+\frac{1}{L}\pi\rho_2'\cot\pi \rho_2
$$
where the $1/L$ correction comes from the contribution to the sum from
the roots separated by $\mathcal{O}(1)$. On the other hand the
auxiliary roots appear as\footnote{recall that the Bethe roots
$u_{2,k}$ belongs to a $\mathcal{C}_{23}$ cut and therefore is always
well separated from $u_{1,j}$ roots which always belong to
$\mathcal{C}_{13}$ cuts.}
$$
\sum_{j}i \log\frac{u_{2,k}-u_{1,j}+i/2}{u_{2,k}-u_{1,j}-i/2}=-\int\limits_{\mathcal{C}_{13}} \frac{\rho_1(y) }{x-y}dy=-\int\limits_{\mathcal{C}_{13}} \frac{\rho_2(y) }{x-y}dy-\int\limits_{\mathcal{C}_{13}} \frac{\rho_1(y)-\rho_2(y)}{x-y}dy
$$
where the last term accounts for the mismatch in densities in cuts
$\mathcal{C}_{13}$ and is clearly also a $\mathcal{O}(1/L)$ effect.
Bellow we will compute this mismatch and find
\beq
\rho_1(x)-\rho_2(x)=\frac{\Delta\cot_{12}}{2\pi i L}=\frac{\cot^+_{21}-\cot^+_{23}}{2\pi i L} \la{deltarho} \,\,\, , \,\,\, x\in \mathcal{C}_{13}
\eeq
where $\Delta f\equiv f(x+i0)-f(x-i0)$ and
\beq
\cot_{ij}\equiv\frac{p'_i-p'_j}{2}\cot\frac{p_i-p_j}{2}\,. \la{cotij}
\eeq
Thus we find, for $x\in  \mathcal{C}_{23}$,
\beqa\la{master}
\frac{1}{x}+2 \int\limits_{\mathcal{C}_{23}}\hspace{-4.5mm}-\hspace{3mm} \frac{\rho_2(y) dy}{x-y} +\int\limits_{\mathcal{C}_{13}} \frac{\rho_2(y) dy}{x-y}=2 \pi n_{23}^A+\phi_2-\phi_3-\frac{1}{L}\[\cot_{23}- \int\limits_{{\cal C}_{13} }\frac{\Delta\cot_{12}}{x-y} \frac{dy}{2\pi i}\]
\eeqa
As explained before, if we apply the duality transformation, cuts
$\mathcal{C}_{23}$ become cuts $\mathcal{C}_{13}$  and vice-versa and,
to leading order, $p_1 \leftrightarrow p_2$. Thus for cuts
$\mathcal{C}_{13}$ we find precisely the same equation (\ref{master})
with $1 \leftrightarrow 2$, so that for $x\in  \mathcal{C}_{13}$
\beqa
\la{master13}
\frac{1}{x}+2 \int\limits_{\mathcal{C}_{13}}\hspace{-4.5mm}-\hspace{3mm} \frac{\rho_2(y) dy}{x-y} +\int\limits_{\mathcal{C}_{23}} \frac{\rho_2(y) dy}{x-y}=2 \pi n_{13}^A+\phi_1-\phi_3-\frac{1}{L}\[\cot_{13}- \int\limits_{{\cal C}_{23} }\frac{\Delta\cot_{12}}{x-y} \frac{dy}{2\pi i}\]
\eeqa
These two equations describing the finite size corrections for the two
types of cuts of the $su(1,2)$ spin chain are the main results of this
section.

In what follows we will derive this result from two different angles.
Namely, we will find this finite size corrections using a Baxter like
formalism based on transfer matrices for this spin chain in several
representations and by exploiting the duality we mentioned in the
previous section. It will become clear that the generalization to other
NBA equations based on higher rank symmetry groups is straightforward.

\subsection{Derivation using the transfer matrices}\la{Ts}
The central object in the study of integrable systems is the
\textit{transfer matrix} $\hat T(u)$. The algebraic Bethe ansatz formalism
has the diagonalization of such objects as main goal and the Bethe
equations appear in the process of diagonalization (see \cite{Faddeev}
and references therein for an introduction to the algebraic Bethe
ansatz). As functions of a spectral parameter $u$ and of the Bethe
roots $u_{a,j}$ these transfer matrices seem to have some poles at
the positions of the Bethe roots. On the other hand they are defined as
a product of $R$ operators which do not have these singularities. This
means that the residues of these apparent poles must vanish. These
analyticity conditions (on the Bethe roots) turn out to be precisely
the Bethe equations, and thus, if we manage to obtain the eigenvalues of
the transfer matrices, we can use this condition of pole cancellation to
obtain the Bethe equations without going through the algebraic Bethe
ansatz procedure, see for example
\cite{Res1,Res2,oldzab,KSZ}. For the $su(1,2)$ spin chain we have the
following transfer matrices in the anti-symmetric representations:
\beqa\la{Ts}
T_{\bx}(u)&=&e^{-i\phi_2}\frac{Q_1(u-\frac{3i}{4})}{Q_1(u+\frac{i}{4})}\frac{Q_2(u+\frac{3i}{4})}{Q_2(u-\frac{i}{4})}\(\frac{u-\frac{5i}{4}}{u-\frac{3i}{4}}\)^L
\\ \nn&+&e^{-i\phi_1}\frac{Q_1(u+\frac{5i}{4})}{Q_1(u+\frac{i}{4})}\(\frac{u-\frac{5i}{4}}{u-\frac{3i}{4}}\)^L +e^{-i\phi_3}\frac{Q_2(u-\frac{5i}{4})}{Q_2(u-\frac{i}{4})} \(\frac{u-\frac{5i}{4}}{u+\frac{i}{4}}\)^L \;\;, \nn
 \\
T_{\bxx}(u)&=&\bar{ T}_{\bx}(\bar
u)\(\frac{u-\frac{5i}{4}}{u+\frac{5i}{4}}\)^L\nn
\;\;,\;\;T_{\bxxx}(u)=\(\frac{u-\frac{5i}{4}}{u+\frac{5i}{4}}\)^L \,.
\eeqa
One can easily see that the Bethe equations do follow from requiring
analyticity of these transfer matrices.

In \cite{GK} it was shown and emphasized that the $TQ$ relations are
the most powerful method to extract finite size corrections to the
scaling limit of Bethe equations.

In this section we will use the transfer matrices presented above along
with the fact that, due to the Bethe equations, they are good
analytical functions of $u$ to find what are the finite size
corrections to this Nested Bethe ansatz. Since for generic (super)
nested Bethe ansatz the transfer matrices in the several
representations are known, this procedure can be easily generalized for
other NBA's.

The key idea to find the finite size corrections to NBA is to use the
transfer matrices in the various representations to define a new set of
quasi-momenta $q_i$ as the solutions of an algebraic equation whose
coefficients are these transfer matrices. For example, to leading
order,
\beqa
\nn&&T_\bx(u) \simeq e^{ip_1}+e^{ip_2}+e^{ip_3}\,,\\
&&\nn T_\bxx(u) \simeq e^{i(p_1+p_2)}+e^{i(p_2+p_3)}+e^{i(p_3+p_1)}\,,\\
\nn&&T_\bxxx(u)\simeq e^{i(p_1+p_2+p_3)}\;,
\eeqa
so that if we define a set of \textit{exact} quasimomenta $q_i$
by\footnote{Exploiting the similarity between this definition equation
and 4.1 in \cite{KSZ} we can easily generalize this algebraic equation
to a more general $su(K|M)$ super group. More precisely we identify
$e^{2\partial_u}\leftrightarrow e^{iq}$ which is obviously natural if
we look at 4.2 in this same paper (see also Appendix A where we use
this two expressions slightly modified to match our normalizations). We
thanks V.Kazakov for pointing this out to us.}
\beq
T_\bxxx(u) - e^{i q} \,T_\bxx(u)\(1-\frac{L}{4u^2}\)  + e^{2i q}
\,T_\bx(u)\(1-\frac{L}{4u^2}\) -e^{3i q} =0\,, \la{Tq}
\eeq
then, to leading order, $q_i\simeq p_i$. Notice however that the
coefficients in this equation have no singularities except some fixed
poles close to $u=0$. Thus, defined in this way, the quasi-momenta
$q_i$ constitute a 4 sheet algebraic surface (modulo $2\pi$
ambiguities) such that
\beq
\sq_i-\sq_j=2\pi n_{ij}^A\;\;,\;\;x\in {\cal C}_{ij} \la{qeq} \,,
\eeq
and, needless to say, this is an \textit{exact} result in $L$, it is
not a classical (scaling limit) leading result like (\ref{scaling})! On
the other hand, the expansion at large $L$ of the above algebraic
equation yields
\beqa
\nn &&q_1=p_1+\frac{1}{2L}(+\cot_{12}+\cot_{13})\\
\nn &&q_2=p_2+\frac{1}{2L}(-\cot_{21}+\cot_{23})\\
\nn &&q_3=p_3+\frac{1}{2L}(-\cot_{31}-\cot_{32})\;,
\eeqa
which follows from the expansion
\beqa
\nn&&T_\bx(u)\(1-\frac{L}{4u^2}\)=e^{ip_1}+e^{ip_2}+e^{ip_3}\\
\nn&&-\frac{1}{4L}\[e^{ip_1}(2p'_1-p'_2-p'_3)+e^{ip_2}(p'_1-p'_3)+e^{ip_3}(p'_1+p'_2-2p'_3)\]
+\O\(\frac{1}{L^2}\)\\
&&\nn T_\bxx(u)\(1-\frac{L}{4u^2}\)=e^{i(p_1+p_2)}+e^{i(p_2+p_3)}+e^{i(p_3+p_1)}\\
\nn&&-\frac{1}{4L}\[e^{i(p_1+p_2)}(p'_1+p'_2-2p'_3)+e^{i(p_1+p_3)}(p'_1-p'_3)+e^{i(p_2+p_3)}(2p'_1-p'_2-p'_3)\]
+\O\(\frac{1}{L^2}\)\;,\\
\nn&&T_\bxxx(u)=e^{i(p_1+p_2+p_3)}+\O\(\frac{1}{L^2}\)\;.
\eeqa
of the several transfer matrices. Then, to the first order in $1/L$ the
exact equation (\ref{qeq}) gives, for the quasi-momenta $p_i$
introduced in (\ref{ps}),
\beqa
\la{p23}\sp_2-\sp_3&=&2\pi n_{23}^A-\frac{1}{L}\cot_{23}\;\;,\;\;x\in {\cal C}_{23}\\
\la{p13}\sp_1-\sp_3&=&2\pi n_{13}^A-\frac{1}{2L}\(\cot_{12}+2\cot_{13}+\cot_{32}\)\;\;,\;\;x\in {\cal C}_{13}
\eeqa
where in \eq{p23} we use the fact that function $\cot_{31}-\cot_{21}$ vanishes
under the slash on the cut ${\cal C}_{23}$ since
\beq
\cot_{ij}^+=\cot_{kj}^-\;\;,\;\;x\in {\cal C}_{ik}\,. \la{updown}
\eeq
Equations (\ref{p23}),(\ref{p13}) are the finite size corrections we
aimed at!

Finally $q_2$ must have no discontinuity at a cut ${\cal C}_{13}$ and
therefore
\beq
\la{Dp_2}\Delta p_2=2\pi i\(\rho_1-\rho_2\)=\frac{1}{L}(\cot_{21}^+-\cot_{23}^+)\;\;,\;\;x\in {\cal C}_{13} \,.
\eeq
Thus, replacing the quasi-momenta $p_i$ by its expressions in terms of
resolvents (\ref{ps}) and relating the density of
\textit{auxiliary roots} $\rho_1$ to that of the \textit{middle node}
roots $\rho_2$ through (\ref{Dp_2}), we recover precisely
(\ref{master}) and (\ref{master13}) as announced.

We would like to stress the efficiency of the $TQ$ relations. We were
able to find the \textit{usual} $\cot$ contributions (coming from the
expansion of the $\log$'s of the Bethe equations when the Bethe roots
are close to each other) plus the mismatch in densities of the
different types of roots making the cuts of stacks using only the fact
that due to Bethe equations the transfer matrices in several
representations were analytical functions of $u$. The computation done
in this way is by far more economical than a brute force expansion of
the Bethe equations.

Finally let us make an important remark. To derive (\ref{master13})
from \eq{p13} one should use
\beq
\cot_{12}=-\frac{1}{2\pi i }\int\limits_{{\cal C}_{13}\cup{\cal C}_{23} }\frac{\Delta\cot_{12}}{x-y}dy \la{cot12}
\eeq
which is clearly a valid relation if $\cot_{12}$ has only branch cuts
as singularities. For generic twists and for small enough cuts
$\mathcal{C}_{13}$ and $\mathcal{C}_{23}$ this is the case. Indeed, in
the absence of Bethe roots we have no cuts at all and thus
$p_1-p_2=\phi_2-\phi_1$. Suppose $\phi_2-\phi_1\neq 2\pi n$. Then, by
continuity, when we slowly open some cuts $\mathcal{C}_{23}$ and
$\mathcal{C}_{13}$ then $p_1-p_2$ will start taking positive values
around $\phi_2-\phi_1$ without ever being zero. Thus, if the cuts are
small enough we will never get poles in $\cot_{12}$. In the next
section we will see that the stacks as described in \cite{BKSZ} only
exist when this assumption of absence of poles is right and are
destroyed when $p_1-p_2$ reaches $2\pi n$.

\subsection{Re-derivation using the bosonic duality in the scaling limit}
\la{dererivationanomaly}
In this section let us re-derive the mismatch formula (\ref{deltarho})
using the bosonic duality (\ref{dual}). Besides the obvious advantage
for what concerns our comprehension of having a second derivation there
are systems for which the Bethe equations are known but the algebraic
formalism behind these equations is still not well developed (this is
the case for example for the $AdS/CFT$ Bethe equations proposed by
Beisert and Staudacher which we will study in section \ref{BSsection}).

Denoting
$$
u_{1,i}=u_{2,i}-\epsilon_i\;\;,\;\;\t u_{1,i}=u_{2,i}-\t\epsilon_i
\,\,\, ,
\,\,\, \epsilon \sim 1
$$
and expanding the bosonic duality (\ref{dual}) in the scaling limit
($L\to \infty$) we get
\beqa
\nn \sin(\tau/2)=\sin\(\frac{1}{2}\(\t G_1-G_1+\tau\)\)
\exp\(
\sum_{i=1}^{K_1}\frac{\epsilon_i}{u-u^1_i}+
\sum_{i=1}^{\t K_1}\frac{\t\epsilon_i}{u-u^1_i}\) \,,
\eeqa
where $\tau=\phi_1-\phi_2$. Taking the logarithm of this equation and
differentiating with respect to $u$ we get
$$
\sum\frac{\epsilon_i}{(u-u^1_i)^2}
+\sum\frac{\t\epsilon_i}{(u-u^1_i)^2}=
\frac{\t G'_1-G'_1}{2L}\cot\frac{\t G_1-G_1+\tau}{2}
$$
where we notice that the left hand side is precisely the difference of
resolvents $G_2-G_1-\t G_1$! Thus we find
$$
G_2-G_1-\t G_1=
\frac{\t G'_1-G'_1}{2L}\cot\frac{\t G_1-G_1+\tau}{2}
\simeq
\frac{G'_2-2G'_1}{2L}\cot\frac{G_2-2G_1+\tau}{2}
=\frac{1}{L}\cot_{12} \,.
$$
Finally, by computing the discontinuity of this expression at the cuts
${\cal C}_{13}$ we will get the \textit{mismatch} of the densities of
the roots in  a cut of stacks\footnote{$\Delta f=f^+-f^-$, so that
$\rho=-\frac{\Delta G}{2\pi i}$}
$$
\rho_1-\rho_2=\frac{\Delta\cot_{12}}{2\pi i L}=\frac{\cot_{21}^+-\cot_{23}^+}{2\pi i L} \,,
$$
which was the gap in the chain of arguments presented in the beginning
of the section \ref{anomalies} and leading to (\ref{master}).

Finally let us show that the bosonic duality amounts to a simple
exchange of Riemann sheets in the scaling limit. Consider for example
$$
\tilde p_1 = -\frac{1}{2x}+\t G_1-\t\phi_1 =
-\frac{1}{2x}+G_2-G_1-\t\phi_1 = p_2
$$
since, as we will see more carefully in the next section, $\t
\phi_{1,2}=\phi_{2,1}$.

\section{1-loop shift}\la{1loope}

In \cite{GV1} we explained how to obtain the spectrum of the
fluctuation energies around any classical string solution using the
algebraic curve by adding a pole to this curve. In particular we
reproduced in this way some previous results \cite{C1,C2,C3,C4} where
the semi-classical quantization around some simple circular string
motions were considered by directly expanding the Metsaev-Tseytlin
action \cite{Metsaev:1998it} around some classical solutions and
quantizing the resulting quadratic action. Using the fact that one
extra pole in the algebraic curve means one quantum fluctuation, we can
compute the leading quantum corrections to the classical energy of the state from
the field theory considerations using the algebraic curve alone, as we mentioned in the introduction. This implies a nontrivial relation between
fluctuations on algebraic curve and finite size corrections in Bethe
ansatz as we will explain in greater detail below. In this section we study this relation on the example of the
$su(1,2)$ spin chain and then in section 6 we extend this to the
super-string case.

As mentioned in the introduction, in the scaling limit $u \sim K \sim L
\gg 1$ we are describing some slow and low energetic spin waves,
\beqa
\nn \E=\sum_{j=1}^{K} \epsilon_j=\sum_{j=1}^{K} \frac{1}{u_{2,j}+1/4} \sim 1/L \,,
\eeqa
around the ferromagnetic vacuum of the theory. In this limit the theory
is well described by a Landau-Lifshistz model which is a field theory
with coupling $1/L$ \cite{LL1,LL2,LL3}. Therefore a very nontrivial
property relating fluctuations and finite size corrections in this NBA
should hold:
\begin{itemize}
\item{Suppose we compute the energy shift $\delta\E_{n}^{ij}$ due to the
 addition of a stack with mode number $n$ uniting sheets $p_i$ and $p_j$
 to a given configuration with some finite cuts $\mathcal{C}$.}
\item{Suppose on the other hand that we compute $1/L$ energy expansion $\E=\E^{(0)}+\frac{1}{L}\E^{(1)}+\dots$ of the configuration with the finite cuts $\mathcal{C}$.}
\end{itemize}
From the field theory point of view the first quantity corresponds to
\textit{one of the fluctuation energies} around a classical solution
parameterized by the configuration with the cuts $\mathcal{C}$ whereas
the second quantity, $\E^{(1)}$, is the \textit{$1$-loop shift}
\cite{Frolov:2002av} around this classical solution with energy
$\E^{(0)}$. This $1$-loop shift, or ground state energy, is given by
the sum of halves of the fluctuation energies \cite{Frolov:2002av}
\beq
\E^{(1)}=\frac{1}{2 }\sum_{n} \sum_{ij} \delta\E_{n}^{ij} \, \la{1loop}
\eeq
In fact for usual (non super-symmetric) field theories this sum is
divergent and needs to be regularized. We will see that (\ref{1loop}) can
be generalized and holds for arbitrary local charges
\beq
{\cal Q}_r^{(1)}=\frac{1}{2 }\sum_{n} \sum_{ij} \delta{\cal
Q}_{r,n}^{ij}\,.
\la{1loopG}
\eeq

Let us stress once more that from the Bethe ansatz point of view these
quantities are computed independently and there is a priori no obvious
reason why such relation between fluctuations and finite size
corrections should hold. In this section we will show that Nested Bethe
Ansatz's describing (super) spin chains with arbitrary rank do indeed
obey such property with some particular regularization procedure (for
the Heisenberg $su(2)$ spin chain a similar treatment was carried in
\cite{BF}). Moreover we will see that the regularization mentioned
above also appears naturally from the Bethe ansatz point of view as
some integrals around the origin.

\subsection{1-loop shift and fluctuations}\la{1loopfluc}
In this section we will understand the interplay between fluctuations
and finite size corrections in NBA's in the scaling limit. For
simplicity we are considering the $su(1,2)$ spin chain described in the
previous sections. General $su(N)$ is considered in Appendix B.

Let us pick the leading order integral equation for the densities of
the Bethe roots in the scaling limit \eq{BAEmn} and perturb it by a
single stack, connecting $p^\ii$ with $p^\jj$. According to
\eq{defdens} this means simply implies
$\rho_2\rightarrow\rho_2+\frac{1}{L}\delta(x-x^{\ii\jj})$, where
$x^{\ii\jj}$ is position of the new stack. Finally, the positions where
one can put an extra stack, as it follows from the BAE
\eqs{eq1}{eq2}, can be parametrized by one integer mod number $n$
\beq
p_\ii(x_n^{\ii\jj})-p_\jj(x_n^{\ii\jj})=2\pi n\,. \la{xnpos}
\eeq
Therefore, for $\ii=2,\;\jj=3$ the perturbed equation \eq{BAEmn} reads
\beq
\la{inteqrho}
\frac{1}{x}+2\sint{{\cal C}_{23}}{4.5mm}\hspace{-3mm} \frac{\rho(y)}{x-y}
+\int\limits_{{\cal C}_{13}} \frac{\rho(y)}{x-y}+\frac{1}{L}\frac{2}{x-x_n^{23}}
=2\pi k_{23}+\phi_2-\phi_3\;\;,\;\;
x\in{\cal C}_{23} \,.
\eeq
and this perturbation will lead to some perturbation of the density
$\delta\rho(y)$, which will lead to the perturbation in the local
charges \eq{su12charges} as
\beq
\delta{\cal Q}_{r,n}^{23} = \int\frac{\delta\rho(y)}{y^r}dy+\frac{1}{L (x_n^{23})^r}\,,
\eeq
the local charges of the fluctuation with polarization ${23}$ and mode
number $n$.

Thus, by linearity, if we want to obtain the $1$-loop shift
(\ref{1loopG}) (or rather a large N regularized version of this
quantity where the sum over $n$ goes from $-N$ to $N$) we have to solve the following integral equation for
densities
\beq
\la{inteqrho}
\frac{1}{x}+2\sint{{\cal C}_{23}}{4.5mm}\hspace{-3mm} \frac{\rho(y)}{x-y}+\int\limits_{{\cal C}_{13}} \frac{\rho(y)}{x-y}
+\sum_{n=-N}^N\frac{1}{2L}\[\frac{2}{x-x_n^{23}}+\frac{1}{x-x_n^{13}}\]=2\pi k_{23}\;\;,\;\;
x\in{\cal C}_{23} \,.
\eeq
and then the $1$-loop shifted charges are given
\beqa
\la{Q0} {\cal Q}_r&=&\int\limits_{{\cal C}_{13}\cup{\cal C}_{23}} \frac{\rho(y)}{y^r}dy
+\sum_{n=-N}^N\frac{1}{2L}\[\frac{1}{(x_n^{23})^r}+\frac{1}{(x_n^{13})^r}\] \, \\
&=&\int\limits_{{\cal C}_{13}\cup{\cal C}_{23}} \frac{\rho(y)}{y^r}dy
+\sum_{n=-N}^N\frac{1}{2L}\[\ccw\oint_{x_n^{23}}\frac{\cot_{23}}{y^r}\frac{dy}{2\pi i}
+\ccw\oint_{x_n^{13}}\frac{\cot_{13}}{y^r}\frac{dy}{2\pi i}\] \,. \la{Q1}
\eeqa
To pass from the first line to the second in the above expression we
use that $\cot_{\ii\jj}$ has poles at $x_n^{\ii\jj}$ with unit residue.
We will now understand how to redefine the density in such a way that
the second term is absorbed into the first one. We start by opening the
contours in (\ref{Q1}) around the excitation points $x_n^{ij}$. These
contours will then end up around the cuts $\mathcal{C}_{kl}$ of the
classical solution and around the origin. We will not consider the
contour around $x=0$  -- this contribution would lead to a
regularization of the divergent sum in r.h.s. of (\ref{1loopG}). We
will analyze it carefully in the super-string case, where it leads to
the Hernandez-Lopez phase factor. Then we get
\beqa
{\cal Q}_r&=&\int\limits_{{\cal C}_{13}\cup{\cal C}_{23}}
\frac{\rho(y)}{y^r}dy
+ \frac{1}{2L}\[\cw\oint_{{\cal C}_{13}}\frac{\cot_{23}}{y^r}\frac{dy}{2\pi i}
+\cw\oint_{{\cal C}_{23}}\frac{\cot_{13}}{y^r}\frac{dy}{2\pi i}\] \la{Q2}
\eeqa
Noting that
\beq
\cot_{ij}^+=\cot_{kj}^-\;\;,\;\;x\in {\cal C}_{ik}\,, \la{updown}
\eeq
where the superscript $+$ ($-$) indicates that $x$ is slightly above
(below) the cut, we can write
\beqa
{\cal Q}_r&=&\int\limits_{{\cal C}_{13}\cup{\cal C}_{23}}
\frac{\rho(y)}{y^r}dy
- \frac{1}{2L}\int\limits_{{\cal C}_{13}\cup{\cal C}_{23}}\frac{\Delta\cot_{12}}{y^r}\frac{dy}{2\pi i} \la{Q3}
\eeqa
so that we see that it is natural to introduce a new density,
``dressed" by the virtual particles,
\beq
\varrho=\rho-\frac{1}{2L}\frac{\Delta \cot_{12}}{2\pi i} \la{derdens}
\eeq
so that the expression for the local charges takes the standard form
\beqa
\nn {\cal Q}_r&=&\int\limits_{{\cal C}_{13}\cup{\cal C}_{23}}
\frac{\varrho(y)}{y^r}dy \,.
\eeqa
 Let us now rewrite
our original integral equation (\ref{inteqrho}) in terms of this
dressed density. We will see that the integral equation we are
constructing for this density by requiring a proper semi-classical
quantization will be precisely the equation (\ref{master}) which is the
finite size corrected integral equation arising from the NBA for the
spin chain! This will thus prove the announced property relating finite
size corrections and $1$-loop shift.
\FIGURE[t]{
    \centering
 \resizebox{120mm}{!}{\includegraphics{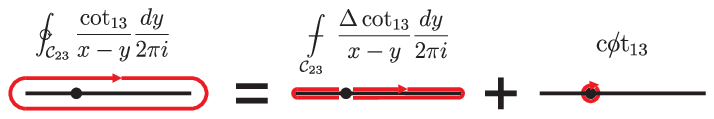}}
        \caption{Illustration of an identity used in the main text.}
        \la{ilust}
 }
Consider for example the first summand in (\ref{inteqrho}) (recall that
$x\in {\cal C}_{23}$),
\beq
\sum_n\frac{1}{x-x_n^{23}}=\sum_n\ccw\oint_{x_n^{23}}\frac{\cot_{23}}{x-y}\frac{dy}{2\pi i}
=\cot_{23}+\cw\oint_{{\cal C}_{13}}\frac{\cot_{23}}{x-y}\frac{dy}{2\pi i}
=\cot_{23}-\int_{{\cal C}_{13}}\frac{\Delta\cot_{12}}{x-y}\frac{dy}{2\pi i} \,, \la{23}
\eeq
Note that $\cot_{23}$ has branch cut singularities at ${\cal C}_{13}$
which we have to encircle when we blow up the contour, which leads to the
second term. The first term comes from the pole at $x=y$. Finally, to
write the second term as it is we used
\eq{updown}. Analogously (see figure
\ref{ilust} for a pictorial explanation of the second equality)
\beq
\sum_n\frac{1}{x-x_n^{13}}
=\cw\oint_{{\cal C}_{23}}\frac{\cot_{13}}{x-y}\frac{dy}{2\pi i}
=\scot_{13}+\sint{{\cal C}_{23}}{4.4mm}\hspace{-3mm}\frac{\Delta\cot_{13}}{x-y}\frac{dy}{2\pi i}
=\scot_{13}-\sint{{\cal C}_{23}}{4.4mm}\hspace{-3mm}\frac{\Delta\cot_{12}}{x-y}\frac{dy}{2\pi i} \,. \la{13}
\eeq
Then we note that (see (\ref{cot12}))
$$
\scot_{13}=\scot_{12}=-\sint{{\cal C}_{13}\cup{\cal C}_{23}}{7.6mm}\hspace{-6mm}\frac{\Delta\cot_{12}}{x-y}\frac{dy}{2\pi i}
$$
so that \eq{inteqrho} reads
$$
\frac{1}{x}+2\sint{{\cal C}_{23}}{4.5mm}\hspace{-3mm} \frac{\rho(y)}{x-y}+\int\limits_{{\cal C}_{13}} \frac{\rho(y)}{x-y}
+\frac{1}{2L}\[2\cot_{23}-2\sint{{\cal C}_{23}}{4.4mm}\hspace{-3mm}\frac{\Delta\cot_{12}}{x-y}\frac{dy}{2\pi i}
-3\int\limits_{{\cal C}_{13}}\frac{\Delta\cot_{12}}{x-y}\frac{dy}{2\pi i}\]=2\pi k_{23}+\phi_2-\phi_3
$$
which in terms of the redefined density $\varrho$ becomes
$$
\frac{1}{x}+ 2\sint{{\cal C}_{23}}{4.5mm}\hspace{-3mm} \frac{\varrho(y)}{x-y}+\int\limits_{{\cal C}_{13}} \frac{\varrho(y)}{x-y}
+\frac{1}{L}\[\cot_{23}
-\int\limits_{{\cal C}_{13}}\frac{\Delta\cot_{12}}{x-y}\frac{dy}{2\pi i}\]=2\pi k_{23}+\phi_2-\phi_3
$$
which coincides precisely with \eq{master} as announced above! Thus the
finite size corrections to the charge of any given configuration will
indeed be equal to the field theoretical prediction, that is to the
$1$-loop shift around the classical solution.

\section{Bosonic duality} \la{dualitysec}
In this section we will explain some details behind the bosonic duality\footnote{Bazhanov and Tsuboi also found some similar duality in the study of the deformed $U_q(sl(1|1))$. We thanks Z.Tsuboi for providing us the talk he gave at the "t9me rencontre entre physiciens theoriciens et mathmaticiens: Supersymmetry and Integrability" (http://www-irma.u-strasbg.fr/article383.html) and V.Kazakov who informed us of their work. It would be very interesting to connect both approaches.}
(\ref{duality}) mentioned in section \ref{one}. There are two main
steps to be considered. On the one hand we have to prove that for a set
of $K_2$ generic complex numbers $u_2$ and $K_1$ roots $u_1$ obeying
the auxiliary Bethe equations \eq{eq1} it is possible to write
($\tau=\phi_1-\phi_2$)
\beq
2i\sin\(\tau/2\) Q_2(u)=e^{i\tau/2}Q_1(u-i/2)\t
Q_1(u+i/2)-e^{-i\tau/2}Q_1(u+i/2)\t Q_1(u-i/2)\,, \la{dual}
\eeq
and that, in doing so, we define the position of a new set of numbers
$\tilde u_1$. A priori this is not at all a trivial statement because
we have a polynomial of degree $K_2$ on the left whereas on the right
hand side we have only $K_2-K_1$ parameters to fix. However, as we will
see, if $K_1$ equations \eq{eq1} are satisfied it is possible to write
$Q_2(u)$ in this form. This will be the subject of the section
\ref{proof}.

Assuming \eq{dual} to be proved we can use this relation to show that
in the original Bethe equations we can replace the roots $u_1$ by the
new roots $\tilde u_1$ with the simultaneous exchange
$\phi_1\leftrightarrow\phi_2$. Indeed if we evaluate the duality at
$u=u_{2,j}$ we find
$$
\frac{Q_1(u_{2,j}-i/2)}{Q_1(u_{2,j}+i/2)}=e^{i(\phi_2-\phi_1)}
\frac{\tilde Q_1(u_{2,j}-i/2)}{\tilde Q_1(u_{2,j}+i/2)
} \,,
$$
meaning that in the equation \eq{eq2} for the $u_2$ roots we can
replace the roots $u_1$ by the dual roots $\t u_1$ provided we replace
$\phi_1\leftrightarrow\phi_2$. Moreover if we take $u=\t u_{1,j}\pm
i/2$ we will get
$$
e^{i\phi_2-i\phi_1}=-\frac{\t Q_1(\t u_1+i)}{\t Q_1(\t
u_1-i)}\frac{Q_2(\t u_1-i/2)}{Q_2(\t u_1+i/2)} \,,
$$
which we recognize as equation \eq{eq1} with $K_2-K_1$ roots $\t u_1$
in place of the $K_1$ original roots $u_1$ and with
$\phi_1\leftrightarrow\phi_2$. Finally evaluating \eq{dual} at
$u=u_{1,j}\pm i/2$ we will get the original equation \eq{eq1} so that we see that it must be satisfied in order to equation \eq{dual} to be valid.

In section \ref{invariance0} we will also see that the transfer
matrices are invariant under the bosonic duality accompanied by an
appropriate reshuffling of the phases $\phi_a$. In section
\ref{examples} some curious examples of dual states will be given.

\subsection{Decomposition proof}\la{proof}

In this section we shall prove that one can always decompose $Q_2(u)$
as in (\ref{dual}) and that this decomposition uniquely fixes the
position of the new set of roots $\t u_1$. In other words, let us show
that we can set the polynomial
\beqa
\nn P(u)\equiv e^{+i\frac\tau2}Q_1(u-i/2)\tilde Q_1(u+i/2)
-e^{-i\frac\tau2}Q_1(u+i/2)\tilde Q_1(u-i/2)
-2i\sin\frac{\tau}{2}Q_2(u)
\eeqa
to zero through a unique choice of the dual roots $\t u_1$.
\begin{itemize}
\item{Consider first the case $K_1=0$. Then it is trivial to see that we
can always find unique
polynomial $\tilde Q_1=u^{K_2}+\sum_{n=1}^{K_2} a_n u^{n-1}$ such that
$$
e^{+i\frac\tau2}\tilde Q_1(u+i/2)
-e^{-i\frac\tau2}\tilde Q_1(u-i/2)=2i\sin\frac{\tau}{2}Q_2(u)\,.
$$
because this amounts to solving $K_2$ linear equations for $K_2$
coefficients $a_n$ with non-degenerate triangular matrix.}
\item{
Next let us consider $K_1\leq K_2/2$. First we choose $\tilde Q_1$ to
satisfy $K_1$ equations
$$
\tilde Q_1(u_p^1)=2 ie^{-i\frac{\tau}{2}} \sin\frac{\tau}{2}\frac{Q_2(u_p^1-i/2)}{Q_1(u_p^1-i)}\equiv c_p
\;\;,\;\;p=1,\dots,K_1
$$
these conditions will define $\tilde Q_1(u)$ up to a homogeneous
solution proportional to $Q_1(u)$,
$$
\tilde Q_1(u)=Q_1(u)\tilde q_1(u)+\sum_{p=1}^{K_1}\frac{Q_1(u)}{Q_1'(u_p^1)(u-u_p^1)}\ c_p
$$
where $\tilde q_1(u)$ is some polynomial of the degree $K_2-2K_1$. Now
from \eq{eq1} we notice that with this choice of $\tilde Q_1$ we have
$$
\frac{P(u_p^1+i/2)}{Q_2(u_p^1+i/2)}=\frac{P(u_p^1-i/2)}{Q_2(u_p^1-i/2)}=0\;\;,\;\;p=1,\dots,K_3
$$
and thus
$$
P(u)=Q_1(u+i/2)Q_1(u-i/2)p(u)
$$
where
$$
p(u)=e^{i\frac{\tau}{2}}\tilde q_1(u+i/2)-e^{-i\frac{\tau}{2}}\tilde
q_1(u-i/2)-2i\sin\frac{\tau}{2}\ q_2(u)
$$
and $q_2$ is a polynomial. Thus we are left to the same problem as
above where $K_1=0$.
For completeness let us note that we can write $q_2(u)$ explicitly in
terms of the original roots $u_1$ and $u_2$,
$$
q_2(u)=\frac{Q_2(u)}{Q_1(u+i/2)Q_1(u-i/2)}-{\rm poles}
$$
where the last term is a simple collection of poles at $u=u_p^1\pm i/2$
whose residues are such that $q_2(u)$ is indeed a polynomial.}

\item{We can see that the number of the solutions
of \eq{eq1} with $K_1=K$ and $K_1=K_2-K$ is the same (see
\cite{Faddeev} for examples of states counting). Thus for each
solution with $K_1\ge K_2/2$ we can always find one dual solution with
$K_1\le K_2/2$ and in this way we prove our statement for $K_1\ge
K_2/2$}

\item{Finally let us stress the uniqueness of the $\tilde Q_1$. If $K_1>\t K_1$
 we have nothing to show since we saw explicitly above how the bosonic duality
  constrains  uniquely the dual polynomial $\t Q_1$. Let us then consider
  $K_1<\t K_1$ and assume we have two different solutions $\tilde Q_1^1$
  and $\tilde Q_1^2$.
Then from the duality relation (\ref{dual}) for either solution we find
\beqa
e^{i\frac{\tau}{2}}Q_1(u-i/2)\(\tilde Q_1^1(u+i/2)-\tilde Q_1^2(u+i/2)\)=\nn\\
e^{-i\frac{\tau}{2}}Q_1(u+i/2)\(\tilde Q_1^1(u-i/2)-\tilde
Q_1^2(u-i/2)\) \nn \,.
\eeqa
Evaluating this expression at $u=u_{1,j}+i/2$ we find that $\t
Q_1^1(u_{1,j})-\t Q_1^2(u_{1,j})=0$ so that $\t Q_1^1(u_1)-\t
Q_1^2(u_1)=Q_1(u) h(u)$ and therefore
$$
e^{i\frac{\tau}{2}} h(u+i/2) =e^{-i\frac{\tau}{2}}h(u-i/2)  \,
$$
which is clearly impossible for polynomial $h(u)$ -- for large $u$ we
can neglect the $i/2$'s to obtain $e^{i\tau}=1$ thus leading to a
contradiction.  }
\end{itemize}
\subsection{Transfer matrix invariance under the bosonic duality}\la{invariance0}
In this section we will examine the transformation properties of the
transfer matrices under the bosonic duality. In Appendix A we consider
this problem for the general $su(N|M)$ group. For now let us just take
$T_{\bx}$ for $su(1,2)$ from \eq{Ts}. Using \eq{dual} we can express
ratios of $Q_1$'s through $\t Q_1$ and $Q_2$ so that
\beqa
\nn T_{\bx}(u)&=&e^{-i\phi_2}\(+\frac{2i\sin\frac\tau2
e^{-i\frac\tau2}Q_2(u-\frac{i}{4})}{Q_1(u+\frac i4)\t Q_1(u+\frac
i4)}+e^{-i\tau}\frac{\t Q_1(u-\frac{3i}{4})}{\t
Q_1(u+\frac{i}{4})}\)\frac{Q_2(u+\frac{3i}{4})}{Q_2(u-\frac{i}{4})}\(\frac{u-\frac{5i}{4}}{u-\frac{3i}{4}}\)^L
\\ \nn &+&e^{-i\phi_1}\(-\frac{2i\sin\frac\tau2
e^{+i\frac\tau2}Q_2(u+\frac{3i}{4})}{Q_1(u+\frac i4)\t Q_1(u+\frac
i4)}+e^{+i\tau}\frac{\t Q_1(u+\frac{5i}{4})}{\t
Q_1(u+\frac{i}{4})}\)\(\frac{u-\frac{5i}{4}}{u-\frac{3i}{4}}\)^L\\
 &+&e^{-i\phi_3}\frac{Q_2(u-\frac{5i}{4})}{Q_2(u-\frac{i}{4})} \(\frac{u-\frac{5i}{4}}{u+\frac{i}{4}}\)^L \;\;. \nn
\eeqa
We see that for $\tau=\phi_1-\phi_2$ the terms with $\sin\frac\tau2$
cancel and we get the old expression for $T_\bx$ with $u_1$ replaced
by $\t u_1$ and $\phi_1\leftrightarrow\phi_2$.

This simple transformation property of the transfer matrices
automatically implies that the Riemann surface defined by the algebraic
equation \eq{Tq} is untouched under the duality transformation (to all
orders in $L$), so that the duality can cause at most some reshuffling of
the sheets. However, as we will see in the next section, not necessarily the sheets as a whole are exchanged -- this operation will be in
general done in a piecewise manner.

\subsection{Examples}\la{examples}

In this section we will study some curious Bethe roots distributions
for the twisted $su(1,2)$ spin chain described by the nested Bethe
equations (\ref{eq1}) and (\ref{eq2}) and for the usual $su(2)$
Heisenberg chain,
\beqa
\(\frac{u_{1,j}+\frac{i}{2} }{u_{1,j}-\frac{i}{2} }\)^L&=&-\frac{Q_1\(u_{1,j}+i \)}{Q_1\(u_{1,j}-i \)} \la{su2}\,.
\eeqa
Using the first example we shall understand the importance of twists to
stabilize big cuts of stacks like the ones depicted in figures $2a$,
$2b$ and explain how the stacks gets destroyed as we decrease the twists.

We can dualize $su(2)$ solutions of the twisted\footnote{For zero twist
the duality becomes degenerate and we will see below that it needs to
be slightly modified.} Heisenberg ring using the same duality
(\ref{duality}) as before with $Q_2 (u)\to u^L$. We will consider the
dual solutions to the vacuum and to a $1$-cut solution for the
Heisenberg spin chain (\ref{su2}) as a prototype of the curious solutions
one would get.

\subsubsection{ Big enough twists, small enough fillings and \textit{zippers} }
In the previous sections we saw that the introduction of twists in the
NBA equations are needed to have a configuration with auxiliary roots $u_1$
close to some momentum carrying roots $u_2$. In figure \ref{f4} we have two
numerical solutions of the Bethe equations which are related by the
bosonic duality. In either of them we see a configuration of Bethe roots
with a simple cut with middle roots only (in blue) and a cut of stacks
(containing blue and yellow roots). In this situation it is clearly
reasonable to think of stacks as bound states of different types of
roots and we see that they indeed condense into \textit{multicolor}
cuts.

\FIGURE{
    \centering
        \resizebox{141mm}{!}{\includegraphics{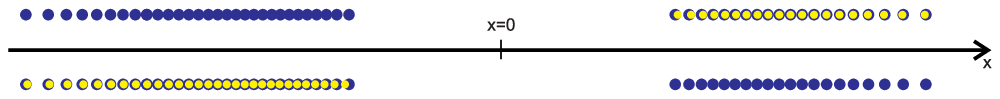}}
    \caption{
    The upper and the lower configuration of Bethe roots
    are dual to one another. Big blue dots are middle node roots $u_2$, yellow dots are
    auxiliary roots $u_1$. The formation of cuts of stacks is
    manifest for this situation where the twists are large (like $\pi/2$)
    and the filling fractions are small.}\la{f4}
}

We will examine what happens when we decrease the twists (or increase
filling fractions, which is the same qualitatively). For simplicity we
consider the configuration, dual to the simple one cut solution
($K_2=K$ and $K_1=0$) with no twist for the middle node roots,
$\phi_2-\phi_3=0$, and some generic twist $\phi_1-\phi_2=\tau$ for the
auxiliary roots. Bosonic duality will leave untouched middle node roots
$u_2$ and create $K$ new axillary roots $u_1$.

\FIGURE{
\centering
        \resizebox{150mm}{!}{\includegraphics{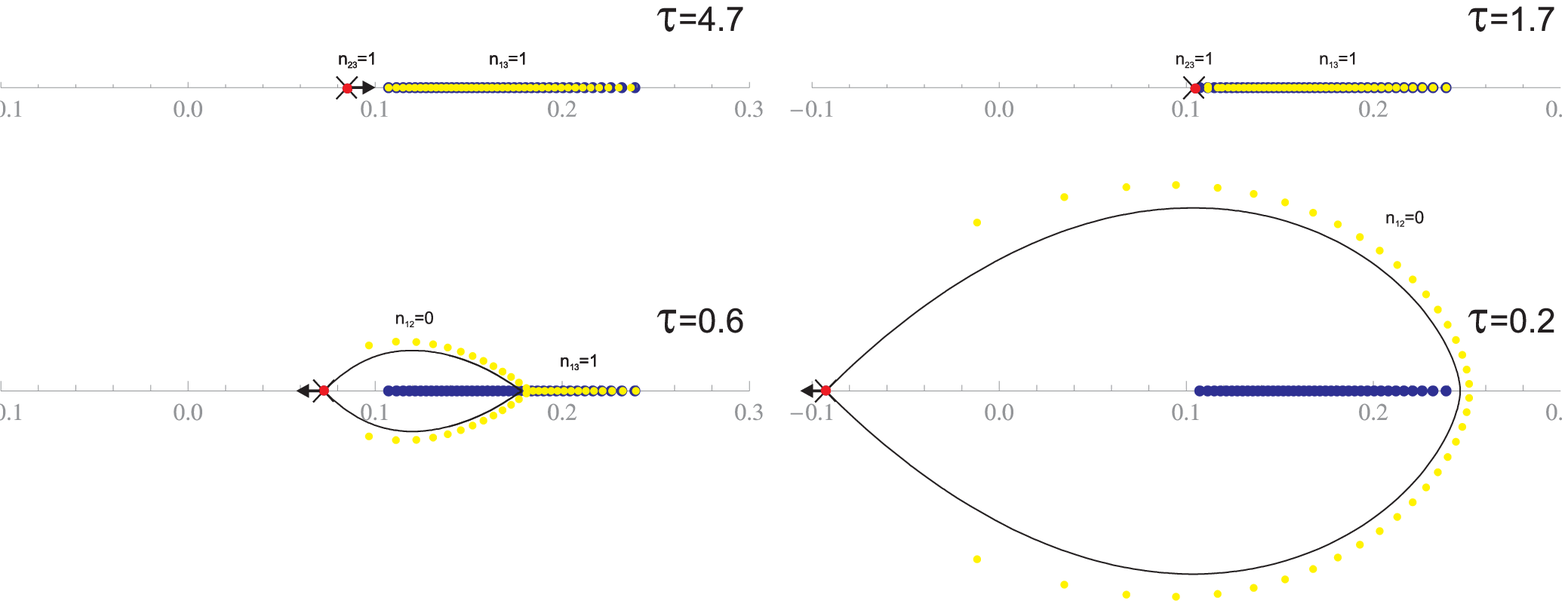}}
    \caption{
    Disintegration of the stack configuration.
    When the twist is large (the top left corner) the
     auxiliary roots form bound states together with the middle
     node ones and constitute a cut of stacks. As
     we decrease the twist fluctuation $n_{23}=1$ (the red crossed dot) enters the cut of
     stacks (the top right corner) and subsequently partly \textit{disintegrate}
     the cut of stacks forming some zipper like configuration (the bottom left corner). At some
     very small value of the twist
     the configuration of Bethe roots bears no resemblance with a cut of stacks.}\la{f5}
}

In the upper left corner of figure
\ref{f5} we applied the duality for some big twist $\tau=4.6$ while in
the bottom right corner of the same figure we have a configuration of
Bethe roots with some small twist $\tau=0.2$. In this latter case the
auxiliary (yellow) roots clearly do \textit{not} form stacks together
with the middle node (blue) roots!, rather they form a bubble,
containing the  original cut of roots $u_2$.

\FIGURE[h]{
    \centering
        \resizebox{70mm}{!}{\includegraphics{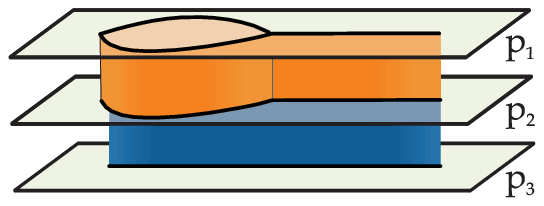}}
    \caption{In the scaling limit the algebraic curves for $e^{ip_j}$
    are the same before the duality (blue cut only)
    and after the duality (when the auxiliary roots are created).
    The duality causes interchange of the sheets outside the bubble,
    while keeping the order untouched inside. This follows from the
    need of a positive density for the ``virtual" cut. In other words the duality
    is indeed only interchanging the sheets of the Riemann surface
    although it is interchanging them in a piecewise way.}\label{zipper}
}

To understand what happens in the scaling limit consider the position
of $n_{23}=1$ fluctuation, given by (\ref{xnpos}), which would be a
small infinitesimal cut between $p_2$ and $p_3$. Clearly this probe cut
would have no influence on the leading order algebraic curve for $p_i$.
In figure \ref{f5} the position of this virtual fluctuation is marked
by a red crossed dot. When the twist is big enough (and filling
fraction is small enough) the fluctuation is to the left from the cut.
When we start decreasing the twist the fluctuation approaches the cut
(upper right picture on fig \ref{f5}) and at this point we have at the
same time
$$
p_2(x_n)-p_3(x_n)=2\pi
$$
and
$$
p_1(x_n)-p_3(x_n)=2\pi\;,
$$
which implies $p_1-p_2=0$ so that equation \eq{cot12} becomes wrong at
this point. When we continue decreasing the twist the fluctuation
passes through the cut and becomes a $n_{12}=0$ fluctuation. If we
think of the fluctuation as being a small cut along the real axis we
see that density becomes negative after crossing the cut:
$$
0<\rho^{fluc}_{23}=-\frac{\Delta(p_2-p_3)}{4\pi
i}=-\frac{\Delta(-p_1-p_2)}{4\pi i}=-\rho^{fluc}_{12}
$$
This means that two branch points of the infinitesimal cut should not
be connected directly, but rather by some macroscopical curve with real
positive density! This curves $z(t)$ can be calculated from the
equation $\rho(z)dz\in {\mathbb R}^+$ or
$$
\frac{p_1(z)-p_2(z)}{2\pi i}\d_t z=\pm 1
$$
and the resulting curve is plotted in black on the two bottom pictures on the
figure \ref{f5}. This is very similar to what happens when a fluctuation
passes through the 1 cut $su(2)$ configuration \cite{BBG}. In the
scaling limit the black curve corresponds to the cut connecting $p_1$
and $p_2$ like on the figure \ref{zipper}.

At first sight these figures seem to be defying our previous results.
Indeed we checked in the previous section that the transfer matrices
themselves are invariant under the bosonic duality. Thus the algebraic
curves obtained from (\ref{Tq}) should be the same after and before
duality and thus what one naturally expects is a simple interchange of
Riemann sheets $p_1\leftrightarrow p_2$ under the duality
transformation. What really happens is a bit more tricky. The
quasimomenta are indeed only exchanged but this exchange operation is
done in a piecewise manner. That is,if we denote the new quasi-momenta
by $p_i^{new}$ and the old ones by $p_i^{old}$ and if we denote the
bubble in figure
\ref{zipper} by $\mathcal{R}$ then we have
$$
p_1^{new}=\left\{\begin{array}{ll}
p_2^{old}  & \textit{, outside\,\,} \mathcal{R} \\
p_1^{old}  & \textit{, inside\,\,} \mathcal{R}
\end{array}\right. \,\, , \,\, p_2^{new}=\left\{\begin{array}{ll}
p_1^{old}  & \textit{, outside\,\,} \mathcal{R} \\
p_2^{old}  & \textit{, inside\,\,}  \mathcal{R}
\end{array}\right. \,\, , \,\,
p_3^{new}=p_3^{old}
$$
where the border of the region $\mathcal{R}$ can be precisely
determined in the scaling limit as explained above.

\subsubsection{ Dualizing momentum carrying roots }\la{examplessu2}
In this section we will consider an example of application of the
bosonic duality to the Heisenberg magnet\footnote{This section
beneficed a lot from the insightful discussions with T.~Bargheer and
N.~Beisert whom we should thank.}. The duality (\ref{duality}) can be
applied to the roots $u_1$ obeying (\ref{su2}) provided we replace
$Q_2(u)\to u^L$. In fact if we want to consider strictly zero twist we
need a new duality because that one is clearly degenerate in this
limiting case. The proper modified expression is in this case
\beqa
\nn&&i(\t K_1-K_1) u^L=Q_1(u-i/2)\t Q_1(u+i/2)- Q_1(u+i/2)\t Q_1(u-i/2)\,.
\la{su2zerotwist}
\eeqa
and the number of dual roots is now $L-K_1+1$. Contrary to what
happened with non-zero twists, here, the dual solution is not unique.
Indeed if $\t K_1> K_1$ we can as well use
\beq
\tilde  Q_1^{\alpha}\equiv \alpha\, Q_1+\t Q_1 \,. \la{dualalpha}
\eeq
\FIGURE{
\centering
        \resizebox{150mm}{!}{\includegraphics{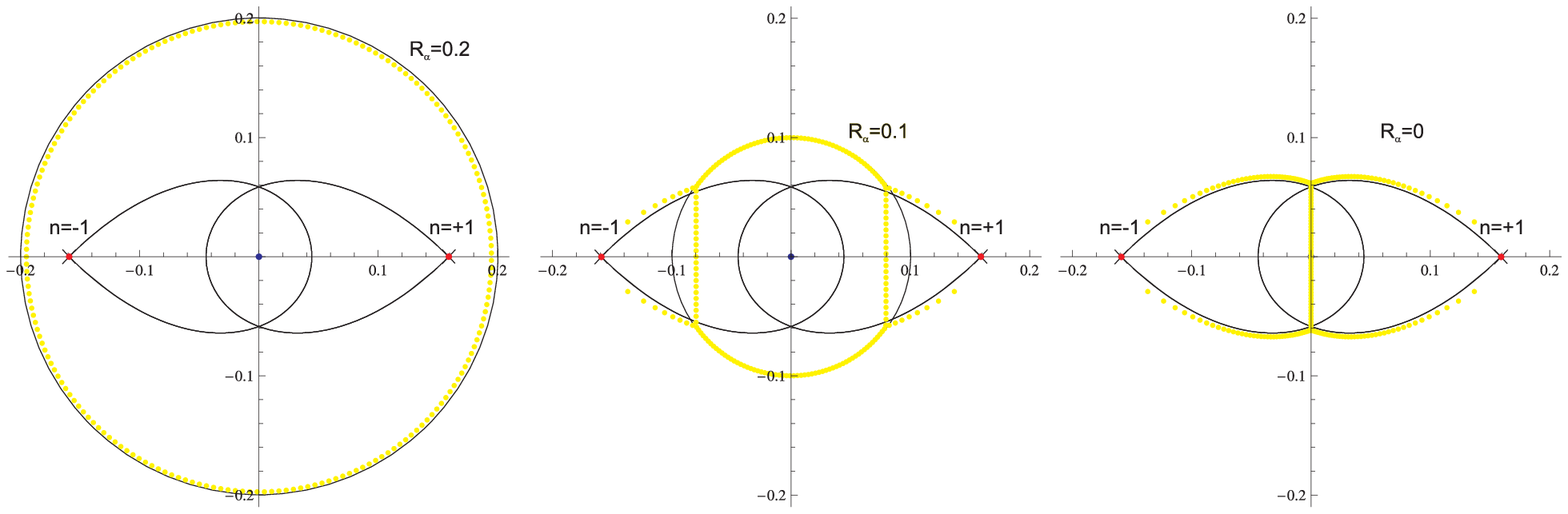}}
    \caption{Three configurations of Bethe roots dual to the ferromagnetic
    vacuum of the untwisted Heisenberg spin chain. For each physical solution (below
    half filling) of the Bethe equations there is a one parameter ($\alpha$)
    family of dual unphysical solutions. To the left, $\alpha$ is large and the roots distribute themselves along a circle with radius $R_{\alpha}$ given by $(R_{\alpha} L)^L= \alpha$.
    Decreasing $\alpha$ the circle will touch the fluctuations $n=\pm 1$. Similarly to the previous section
    the virtual infinitesimal cuts become macroscopical bubble cuts with cusps at the position of the fluctuations. Intersection points of the new cuts with the circle are connected by condensates, which are logarithmic cuts on the algebraic curve \cite{BBG}.}\label{f7} }
All these solutions, parameterized by the constant $\alpha$, have the
same charges because the transfer matrix is invariant under this
transformation -- see appendix A. Notice that if initially we have a
physical state with $K_1<L/2$ roots then all dual states
(\ref{dualalpha}) are unphysical with $\t K_1>L/2$ violating the
half-filling condition. Still, it is interesting, at the level of Bethe
equations, to understand how these solutions look like. First of all
let us single out a particular $\t Q_1$ out of the various solutions to
(\ref{su2zerotwist}) so that
\beq
\tilde Q_1^{\alpha} = u^{\t K_1}+\sum_{l=0}^{\t K_1-1} c_l^{\alpha} \, u^l
\eeq
becomes well defined through (\ref{dualalpha}). We chose $\t Q_1=\tilde
Q_1^0$ to be the dual solution with $c^{0}_0=0$.

Consider for example the vacuum state for which $Q_1=1$. Let us first
take $\alpha$ to be very large so that we can write
\beq
\alpha+\tilde Q_1^0\simeq \alpha+(x L)^L\,.
\eeq
We see for large $\alpha$ the dual roots will be on a circle of radius
$\frac{|\alpha|^{1/L}}{L}$. The corresponding configuration is present
on the first picture on the figure \ref{f7}. In this figure we also plotted a circle with this radius and one can see that the
Bethe roots belong perfectly to the circle.

Let us now understand this configuration from the algebraic curve
point of view. The the quasi-momenta $p_1=-p_2\equiv p=\frac{1}{2x}-G$,
in the absence of Bethe roots, are simply given by $p=\frac{1}{2x}$.
Let us find the curves with positive densities and mode number $n=0$. The density is given by
$\rho(x)=\frac{1}{2\pi i}\frac{1}{x}$ and we have to find the curves
where $\rho(x)dx$ is real. It is easy to see that the only possibility
is the circle centered at the origin with an arbitrary radius. From the
above arguments one can expect that for any $\alpha$ the roots will
belong to some circle. However, we analysed only the curves with zero mode number and as we see on the figure
\ref{f7} for smaller $\alpha$'s the circle develops four tails and two vertical lines. Along these vertical lines the roots are separated by $i$ (for $L\to
\infty$) forming the so called \textit{condensates} or \textit{Bethe strings}. The tails meet at the points where the virtual fluctuation is and the corresponding curves are given by
\beq
\frac{p(z)\pm \pi }{\pi i}\d_t z=\pm 1\;
\eeq
analogously to the previous section. In the last configuration on figure \ref{f7} the circle is completely
absent. There are only two $n=\pm 1$ curves which, at the
interceptions, become a $4\pi$ jump log condensate with the Bethe roots separated by $i/2$.

\FIGURE{
\centering
        \resizebox{150mm}{!}{\includegraphics{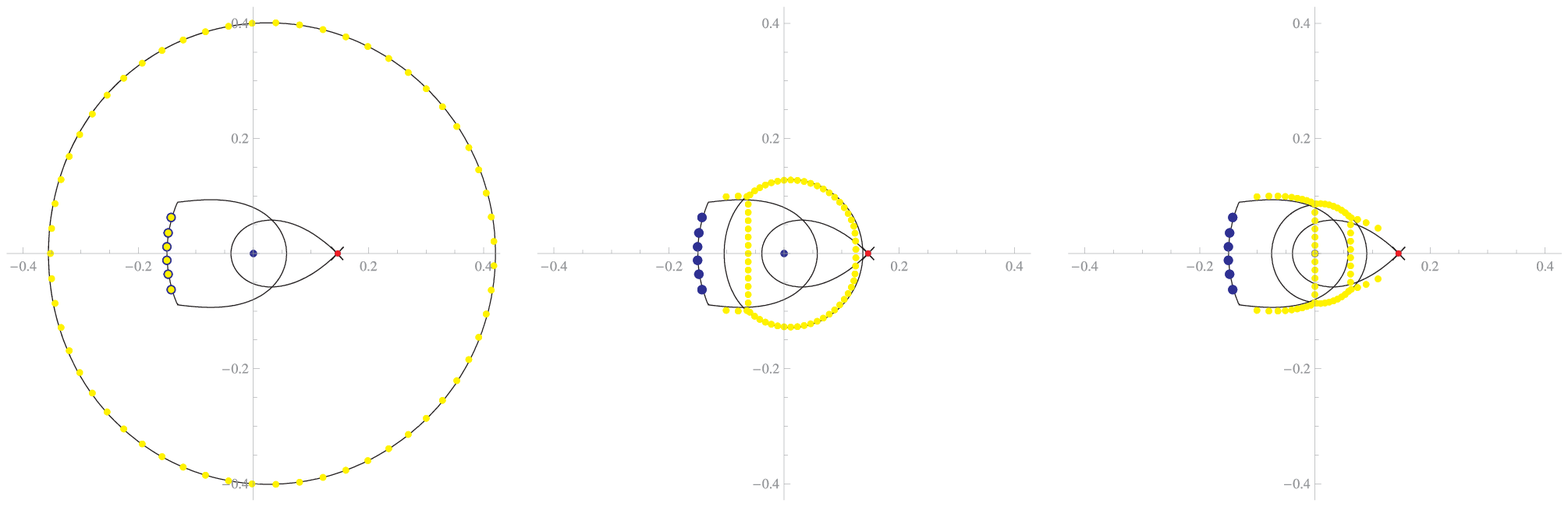}}
    \caption{Dual configuration to 1-cut solution. Similar to the previous example
    for the large $\alpha$ the dual roots are distributed along the big circle and cut (first picture).
    When the $\alpha$ decreases and the circle crosses the cut we have to choose another curve with the positive density (second and third pictures).}\label{f1cut}
 }
We also built the dual configurations to the 1-cut solution (see figure
\ref{f1cut}). The situation is similar to the vacuum, the only difference being that
two tails (out of four) do not tend to touch each other, but rather end at the branch points of the initial cut.

\section{The AdS/CFT Bethe equations and the semiclassical
quantization of the superstring on $AdS_5\times S^5$}\la{BSsection}
\subsection{Introduction and notation}
The Beisert-Staudacher (BS) equations \cite{BS} are a set of $7$ asymptotic \cite{r56} Bethe 
equations (the rank of the symmetry group $PSU(2,2|4)$) which are
expected to describe the anomalous dimensions of $\mathcal{N}=4$ SYM
single trace operators with a large number of fields\footnote{These
large traces can be though of as spin chains and then the dilatation
operator behaves like a spin chain Hamiltonian which turns out to be
integrable \cite{MZ,B}. In this way Bethe equations appear naturally
from the gauge theory side.} as well as the energy of the dual string
states\footnote{The existence of a finite gap description of the
classical string motion \cite{KMMZ,BKSZ} lead to the belief that these
equations ought to be the continuous limit of some quantum string Bethe
equations. In other words, the Riemann surfaces present therein should
in fact be the condensation of a large number of Bethe roots. Inspired
by these finite gap constructions these quantum equations were proposed
shortly after \cite{AFS,BS}.}. The perturbative gauge theory and the
classical string regimes are interpolated by these equations through
the t'Hooft coupling $\lambda$. In \cite{Beisert:2005tm}, based on an
hypothesis for a natural extension for the quantum symmetry of the
theory, Beisert found (up to a scalar factor) an S-matrix from which
the BS equations would be derived. The scalar factor was then
conjectured in \cite{Beisert:2006ib,Beisert:2006zy} from the string
side -- using the Janik's crossing relation \cite{Janik:2006dc} -- and
in \cite{Beisert:2006ez,Eden:2006rx} from the gauge theory point of
view -- based on several heuristic considerations
\cite{Kotikov:2002ab}. From the gauge theory side these
equations were tested quite recently up to four loops
\cite{test1,test2,test3}. From the string theory point of view the
scalar factor recently passed several nontrivial checks
\cite{Dorey:2007xn,J2,Kro,TR} where several loops were probed at strong
coupling. Also at strong coupling, the full structure of the BS equations was derived up to two loops
in \cite{Z2,Z3} in a particular limit \cite{Maldacena:2006rv} where the sigma model is drastically simplified.

In this section we will check that the BS equations reproduce the
$1$--loop shift around \textit{any} classical string soliton solution with
exponential precision in the large angular momentum in the string
state. To do so our computation is divided into two main steps. On the
one hand we will compute the $1/\sqrt{\lambda}$ corrections to
Bethe equations in the scaling limit. We will have to use the
technology developed in the previous sections in order to understand
precisely the several sources of corrections, the most subtle of all
being the fine structure of the cuts of stacks which are generically
present\footnote{In \cite{r73,r74,r75} the scaling limit of the $SU(3)$ sector was considered. It would be interesting to use our treatment, including stacks, to compute explicitly the finite size corrections in this subsector following the lines of these papers.}. At the end we will find out some integral equation
corrected by a $1/\sqrt{\lambda}$ term.

On the other hand we start from the algebraic curve description of the
string classical motion \cite{KMMZ,BKSZ}. The integral equations
present in this finite gap formalism coincide with the scaling limit of
the Bethe equations. Then we find how to correct this equations in such
a way that they will now describe not only the classical motion but
also the semi-classical quantization of the theory around \textit{any
classical motion}. For example we will find out how to modify the
equations in such a way that they exhibit a very nontrivial property:
the first finite corrections to any classical configurations
equals the sum of quantum fluctuations around this same classical
configuration. Then we show that, modified  in this way, the integral
equations coincide precisely with the scaling limit expansion of the BS equations with the HL phase \cite{HL} (up to some exponencially supressed wrapping effects, irrelevant for
 large angular momentum string states)! In this way we establish that, to this order in
$1/\sqrt{\lambda}$, the BS equations do provide the correct
quantization of the system.

These Bethe equations are a deformation of the equations
(\ref{BAEproduct}) through the introduction of the map
$$
x+\frac{1}{x}=\frac{4\pi
u}{\sqrt\lambda}\;\;,\;\;x^\pm+\frac{1}{x^\pm}=\frac{4\pi}{\sqrt\lambda}\(u\pm\frac
i2\) \,.
$$
As explained in section \ref{one} for superalgebras the choice of
Bethe equations is not unique. In \cite{BS} four choices are presented.
We need only to consider two of them\footnote{In \cite{BS} we consider
$\eta_1=\eta_2=\eta$.}, corresponding to the diagram in figure
\ref{figH0} or to the reflected path along the diagonal going from the
lower left to the upper right corner. 

Moreover we consider a twisted
version of these equations for the same reasons mentioned in the
previous sections. In \cite{Beisert:2005if,Frolov:2005iq} a similar kind of twists were introduced in the study of a set of deformations of
$\mathcal{N}=4$ SYM and of the dual sigma model. 
Our twists seem to be a simple change in boundary conditions via the introduction of a constant matrix like (\ref{g}). It would be interesting to see if they can also be given a deeper physical interpretation following the lines of these
works. We should stress that the twists are used here as a technical tool which will simplify our analysis because, in particular, it allows us to deal with well defined stacks in a regime where the dualities are nothing but an exchange of Riemann sheets. We will explain in section \ref{zerotw} that we can then safely analytically continue the results to zero twist.
\newpage

The BS equations then read
\begin{eqnarray}
\nn e^{i\eta\phi_1-i\eta\phi_2}&=&
\prod_{j=1}^{K_2}
\frac{u_{1,k}-u_{2,j}+\frac{i}{2} }{u_{1,k}-u_{2,j}-\frac{i}{2} }
\prod_{j=1}^{K_4}
\frac{1-1/x_{1,k}x_{4,j}^{+ }}{1-1/x_{1,k}x_{4,j}^{- }}\,,
\\
\nn e^{i\eta\phi_2-i\eta\phi_3}&=&
\prod_{j\neq k}^{K_2}
\frac{u_{2,k}-u_{2,j}-i }{u_{2,k}-u_{2,j}+i }
\prod_{j=1}^{K_3}
\frac{u_{2,k}-u_{3,j}+\frac{i}{2} }{u_{2,k}-u_{3,j}-\frac{i}{2} }
\prod_{j=1}^{K_1}
\frac{u_{2,k}-u_{1,j}+\frac{i}{2} }{u_{2,k}-u_{1,j}-\frac{i}{2} }\,,
\\
\nn e^{i\eta\phi_3-i\eta\phi_4}&=&
\prod_{j=1}^{K_2}
\frac{u_{3,k}-u_{2,j}+\frac{i}{2} }{u_{3,k}-u_{2,j}-\frac{i}{2} }
\prod_{j=1}^{K_4}
\frac{x_{3,k}-x_{4,j}^{+ }}{x_{3,k}-x_{4,j}^{- }}\,,
\\
\la{middle} e^{i\eta\phi_4-i\eta\phi_5}&=&\(\frac{x^-_{4,k}}{x^+_{4,k}}\)^{\eta
L}
\prod_{j\neq k}^{K_4}
\frac{u_{4,k}-u_{4,j}+i}{u_{4,k}-u_{4,j}-i} \,
\prod_{j}^{K_4}
\(\frac{1-1/x_{4,k}^+ x_{4,j}^-}{1-1/x_{4,k}^- x_{4,j}^+}\)^{\eta-1}\,
\(\sigma^2(x_{4,k},x_{4,j})\)^{\eta} \\
\nn && \qquad
\times
\prod_{j=1}^{K_1}
\frac{1-1/x_{4,k}^{- } x_{1,j}}{1-1/x_{4,k}^{+ }x_{1,j}}
\prod_{j=1}^{K_3}
\frac{x_{4,k}^{- }-x_{3,j}}{x_{4,k}^{+ }-x_{3,j}} \prod_{j=1}^{K_5}
\frac{x_{4,k}^{- }-x_{5,j}}{x_{4,k}^{+ }-x_{5,j}}
\prod_{j=1}^{K_7}
\frac{1-1/x_{4,k}^{- }x_{7,j}}{1-1/x_{4,k}^{+ }x_{7,j}}\,,
\\
\nn e^{i\eta\phi_5-i\eta\phi_6}&=&
\prod_{j=1}^{K_6}
\frac{u_{5,k}-u_{6,j}+\frac{i}{2} }{u_{5,k}-u_{6,j}-\frac{i}{2} }
\prod_{j=1}^{K_4}
\frac{x_{5,k}-x_{4,j}^{+ }}{x_{5,k}-x_{4,j}^{- }}\,,
\\
\nn e^{i\eta\phi_6-i\eta\phi_7}&=&
\prod_{j\neq k}^{K_6}
\frac{u_{6,k}-u_{6,j}-i }{u_{6,k}-u_{6,j}+i }
\prod_{j=1}^{K_5}
\frac{u_{6,k}-u_{5,j}+\frac{i}{2} }{u_{6,k}-u_{5,j}-\frac{i}{2} }
\prod_{j=1}^{K_7}
\frac{u_{6,k}-u_{7,j}+\frac{i}{2} }{u_{6,k}-u_{7,j}-\frac{i}{2} }\,,
\\
\nn e^{i\eta\phi_7-i\eta\phi_8}&=&
\prod_{j=1}^{K_6}
\frac{u_{7,k}-u_{6,j}+\frac{i}{2} }{u_{7,k}-u_{6,j}-\frac{i}{2} }
\prod_{j=1}^{K_4}
\frac{1-1/x_{7,k}x_{4,j}^{+ }}{1-1/x_{7,k}x_{4,j}^{- }}\,.
\end{eqnarray}
In fact, in order for the fermionic duality
\cite{BS} (which we will review below) to exist, the twists must not be completely independent but rather
\beqa
\phi_1-\phi_2+\eta\sum_{j=1}^{K_4}\frac{1}{i}\log\frac{x_4^{+}}{x_4^{-}}&=&\phi_3-\phi_4\,\,
, \,\, \nn \\
 \phi_7-\phi_8+\eta\sum_{j=1}^{K_4}\frac{1}{i}\log\frac{x_4^{+}}{x_4^{-}}&=&\phi_5-\phi_6\,.
\la{restriction}
 \eeqa
The energy (the anomalous dimension) can then be read from
\beqa
\delta D=\frac{\sqrt{\lambda}}{2\pi}\sum_{j=1}^{K_4} \(
\frac{i}{x_{4,j}^+}-\frac{i}{x_{4,j}^-} \) . \la{deltaD}
\eeqa
To describe classical solutions (and to semi-classically quantize them)
we should consider the scaling limit where
$$
\sqrt{\lambda}\sim u \sim K_a \sim L \gg 1 \,.
$$
In this limit we have
$$
x^{\pm}=x\pm \frac{i}{2}\,\alpha(x)+\O\(\frac1\lambda\)
$$
where
$$
\alpha(x)\equiv \frac{4\pi}{\sqrt\lambda}\frac{x^2}{x^2-1} \,.
$$
It is then useful to introduce the resolvents\footnote{note that
$$
F_a(x)=G_a(x)+\bar G_a(x)=H_a(x)+\bar H_a(x) \,.
$$}
\beqa
&&\nn F_a(x)=\sum_j\frac{1}{u-u_{a,j}} \,,\\
&&\nn G_a(x)=\sum_j\frac{\alpha(x_{a,j})}{x-x_{a,j}}\;\;,\;\;\bar
G_a(x)=\sum_j\frac{\alpha(1/x_{a,j})}{x-1/x_{a,j}}\\
&&\nn H_a(x)=\sum_j\frac{\alpha(x)}{x-x_{a,j}}\;\;,\;\;\bar
H_a(x)=\sum_j\frac{\alpha(1/x)}{1/x-x_{a,j}}
\eeqa
and build with them eight quasi-momenta ($\J=L/\sqrt{\lambda}$)
\beqa
\bea{l}
p_1 = + \displaystyle{\frac{2\pi \J x  - \delta_{\eta,+1}
\mathcal{Q}_1+
\delta_{\eta,-1}\mathcal{Q}_2 x}{x^2-1}} +\eta\( - H_1 - \bar H_3 + \bar
H_4\)+\phi_1\\
\midrule
p_2 = + \displaystyle{\frac{2\pi \J x - \delta_{\eta,-1}\mathcal{Q}_1+
\delta_{\eta,+1}\mathcal{Q}_2 x }{x^2-1}} +\eta\(- H_1 + H_2 + \bar
H_2 - \bar H_3\)+\phi_2\\
p_3 = +\displaystyle{\frac{2\pi \J x  - \delta_{\eta,-1}\mathcal{Q}_1+
\delta_{\eta,+1}\mathcal{Q}_2 x }{x^2-1}} +\eta\( - H_2 + H_3 + \bar H_1 -
\bar H_2\)+\phi_3\\  \midrule
p_4 = +\displaystyle{\frac{2\pi \J x  - \delta_{\eta,+1}\mathcal{Q}_1+
\delta_{\eta,-1}\mathcal{Q}_2x }{x^2-1}} +\eta\( + H_3 - H_4 + \bar
H_1\)+\phi_4\\
p_5 = -\displaystyle{\frac{2\pi \J x - \delta_{\eta,+1}\mathcal{Q}_1+
\delta_{\eta,-1}\mathcal{Q}_2x }{x^2-1}} +\eta\( - H_5 + H_4 - \bar
H_7\)+\phi_5\\  \midrule p_6 = -\displaystyle{\frac{2\pi \J x -
\delta_{\eta,-1}\mathcal{Q}_1+\delta_{\eta,+1}\mathcal{Q}_2 x }{x^2-1}}
 +\eta\(- H_5 + H_6 + \bar H_6 - \bar
H_7\)+\phi_6\\
p_7 = -\displaystyle{\frac{2\pi \J x   - \delta_{\eta,-1}\mathcal{Q}_1+
\delta_{\eta,+1}\mathcal{Q}_2 x }{x^2-1}} +\eta\( - H_6 + H_7 + \bar H_5 -
\bar H_6\)+\phi_7\\  \midrule
p_8 = -\displaystyle{\frac{2\pi \J x -
\delta_{\eta,+1}\mathcal{Q}_1+
\delta_{\eta,-1}\mathcal{Q}_2 x }{x^2-1}} +\eta\( + H_7 +
\bar H_5 - \bar H_4\)+\phi_8
\eea
\la{eq:p}
\eeqa
where $G_4(x)\equiv -\sum_{n=0}^{\infty} {\cal Q}_{n+1} x^{n}$. We can
also write
$$
\frac{2\pi}{\sqrt{\lambda}} \,\delta {\cal D}=\mathcal{Q}_2 \,.
$$
Then, to leading order, these quasi-momenta define an eight-sheet
Riemann surface and the BS equations read simply $\sp_i-\sp_j=2\pi
n_{ij}$ in each of the cuts $\mathcal{C}_{ij}$ uniting $p_i$ and $p_j$.
Finally, in this section we will use
$$
\cot_{\ii\jj}\equiv\alpha(x)\frac{p'_\ii-p'_\jj}{2}\cot\frac{p_\ii-p_\jj}{2}
$$
which is similar (but should not be confused) with \eq{cotij}.

\FIGURE[t]{
   \centering
       \resizebox{100mm}{!}{\includegraphics{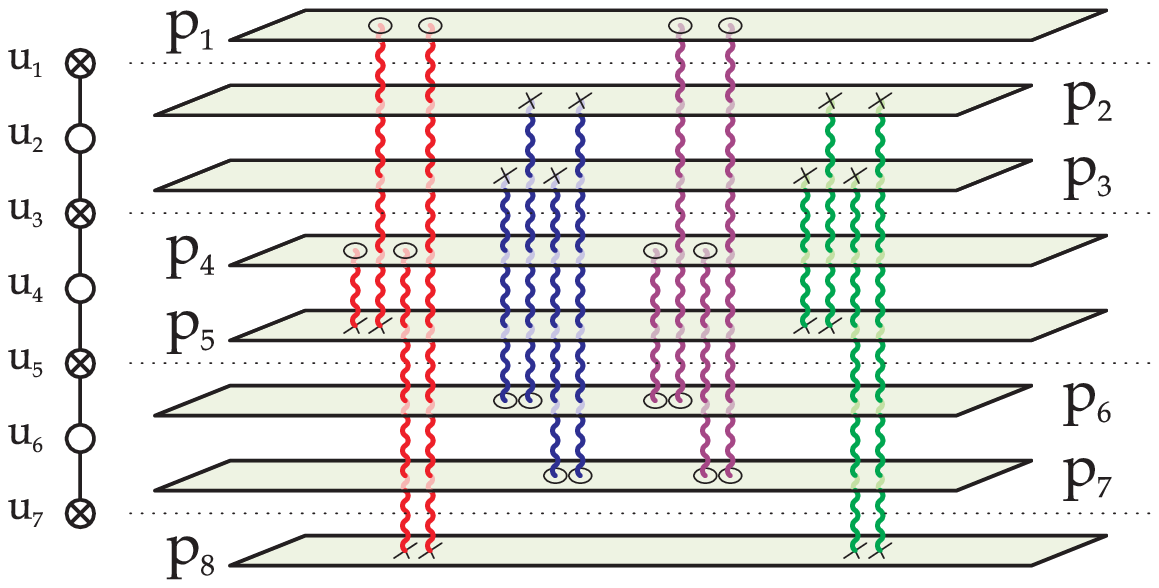}}
   \caption{
The several physical fluctuations in the string Bethe ansatz. The 16
elementary physical excitations are the stacks (bound states)
containing the middle node root. From the left to the right we have
four $S^5$ fluctuations, four $AdS_5$ modes and eight fermionic
excitations. The bosonic (fermionic) stacks contain an even (odd)
number of fermionic roots represented by a cross in the $psu(2,2|4)$ Dynkin diagram in the left.\label{fig:simple} } }

\subsection{Middle node anomaly}
In this section we will expand BS equations in the scaling limit for
the roots belonging to a cut containing middle node roots $x_4$ only. We do not assume that all the others cuts are
of the same type, rather they can be cuts of stacks of several sizes. In the section \ref{examples} we will generalize the results obtained in
this section to an arbitrary cut, assuming, as in the previous
section, that the cuts are small enough and twists are not zero so that stacks are stable. We will discus in section~\ref{zerotw} what happens when
one takes all twists to zero.

To leading order, the middle node
equation (\ref{middle}) can be simply written as $\sp_4-\sp_5=2\pi n$
while at $1$--loop the first product in the r.h.s. of (\ref{middle})
corrects this equation due to
\beqa
\la{eq:mn}&&\frac{1}{i}\log\prod_{j\neq k}^{K_4}
\(\frac{u_{4,k}-u_{4,j}+i}{u_{4,k}-u_{4,j}-i}\)\simeq
 2 \,\,\sF_4(x)+ \alpha(x)\pi\rho'(x)\cot(\pi\rho(x))
\eeqa
where $\rho(x)=\frac{dk}{du_k}$. Expansion of the remaining terms in
\eq{middle} will not lead to the appearance of such \textit{anomaly} like terms since the roots
of another types are separated by $\sim 1$ from $x_{4,k}$. Thus we have simply
\beq\la{midan}
2\nn\pi n=\sp_4-\sp_5
-\eta\,\alpha(x)\pi\rho'(x)\cot(\pi\rho(x))\;\;,\;\;x\in{\cal C}_{45}
\eeq
In the next sections we will use dualities of the BS equations to get
some extra information about cuts of stacks and generalize the above
equation to any possible type of cut. To achieve this we shall recast this equation in terms of the middle node roots $x_4$ only.

%
\subsection{Dualities in the string Bethe ansatz}
Obviously, the behavior of the Bethe roots will be as described in
section \ref{one} for a simpler example of a $su(1,2)$ spin chain, that
is, we will have simple cuts made out of $x_4$ roots only and also cuts of
stacks with $x_2, x_3$ and $x_4$ roots for example. Consider such cut of
stacks. Clearly, to be able to write the middle node equation
(\ref{middle}) or (\ref{midan}) we need to compute the density
mismatches $\rho_2-\rho_3$ and $\rho_3-\rho_4$ which are $1$-loop
contributions we must take into account if we want to write an
integral equation for the middle node equation in terms
of the density $\rho_4$ of momentum carrying roots only. In this
section we shall analyze the dualities present in the BS Bethe
equations. By analyzing them in the scaling limit we will then be able
to derive the desired density mismatches.

\subsubsection{Fermionic duality in scaling limit}
In \cite{BS} it was shown that the BS equations obey a very important
fermionic duality. Since we chose to work with a subset of the possible
Bethe equations, that is the ones with $\eta_1=\eta_2=\eta$ present in
\cite{BS}, we should apply the duality present below not only to the
fermionic roots $x_1$ and $x_3$ (as described below) but also to the Bethe
roots $x_5$ and $x_7$. Obviously the duality for $x_5$ and $x_7$ is
exactly the same as for $x_1$ and $x_3$ and so we will focus simply on
the latter while keeping implicit that we always dualize all the
fermionic roots at the same time.

We construct the polynomial ($\tau=\eta\(\phi_4-\phi_3\)$)
\beqa
\nn P(x)&=&e^{+i\frac\tau2}\prod_{j=1}^{K_4}(x-x_{4,j}^+)\prod_{j=1}^{K_2}(x-x_{2,j}^-)(x-1/x_{2,j}^-)\\
&-&e^{-i\frac\tau2}\prod_{j=1}^{K_4}(x-x_{4,j}^-)
\prod_{j=1}^{K_2}(x-x_{2,j}^+)(x-1/x_{2,j}^+)
\la{eq:fermionic1}
\eeqa
of degree $K_4+2K_2$ which clearly admits $x=x_{3,j}$ and$x=1/x_{1,j}$ as $K_3+K_1$ zeros\footnote{we
also have $1/x_1$ has zeros because, due to (\ref{restriction}), the
equation for $x_{1,j}$ is the same as the equation for $x_{3,j}$ if we
replace $x_{3,j}$ by $1/x_{1,j}$. This is why the restriction (\ref{restriction}) of the twists is so important.} .
The remaining $K_4+2K_2-K_3-K_1$ roots are denoted by  $\t x_{3,j}$ or
$1/\t x_{1,j}$ depending on whether they are outside or inside the unit
circle respectively,
\beq
P(x)=2i\sin(\tau/2)\prod_{j=1}^{K_1}(x-1/x_{1,j})\prod_{j=1}^{\t
K_1}(x-1/\tilde x_{1,j})\prod_{j=1}^{ K_3}(x-x_{3,j})\prod_{j=1}^{\t
K_3}(x-\tilde x_{3,j})\la{eq:fermionic2}
\eeq
Then we can replace the roots $x_{1,j},x_{3,j}$ by the roots  $\t x_{1,j},\t
x_{3,j}$ in the BS equations provided we change the grading $\eta \to
-\eta$ and interchange the twists $\phi_1\leftrightarrow
\phi_2$ and $\phi_3\leftrightarrow \phi_4$. In fact, since we should also dualize the remaining fermionic roots, we should
also change $\phi_5\leftrightarrow \phi_6$ and $\phi_7\leftrightarrow
\phi_8$ and replace the remaining fermionic roots $x_5$ and $x_7$.

Since to the leading order $x^\pm \simeq x$ each root will
belong to a stack which must always contain a momentum carrying root
$x_4$. We have therefore $\t K_1=K_2-K_1$ and $\t K_3=K_2+K_4-K_3$.
Thus we label the Bethe roots as
\beqa
\nn&& x_{1,j}=x_{4,j}-\epsilon_{1,j}\;\;,\;\;\;\;\;\;\;\,j=1,\dots,K_1\\
\nn&& \t x_{1,j}=x_{4,j+K_1}-\t\epsilon_{1,j}\;\;,\;\;j=1,\dots,\t K_1\\
\nn&& x_{2,j}=x_{4,j}-\epsilon_{2,j}\;\;,\;\;\;\;\;\;\;\,j=1,\dots,K_2\\
\nn&& x_{3,j}=x_{4,j}-\epsilon_{3,j}\;\;,\;\;\;\;\;\;\;\,j=1,\dots,K_3\\
\nn&& \t x_{3,j}=x_{4,j+K_3}-\t\epsilon_{3,j}\;\;,\;\;j=1,\dots,\t K_3
\eeqa
with $\epsilon\sim 1/\sqrt\lambda $. Dividing \eq{eq:fermionic1} and
\eq{eq:fermionic2} by
$\prod_{j=1}^{K_4}(x-x_{4,j})\prod_{j=1}^{K_2}(x-x_{4,j})(x-1/x_{4,j})$
we have
\beqa
\nn&&e^{+i\frac\tau2}\prod_{j=1}^{K_4}\frac{x-x_{4,j}^+}{x-x_{4,j}}\prod_{j=1}^{K_2}\frac{x-x_{2,j}^-}{x-x_{4,j}}\frac{x-1/x_{2,j}^-}{x-1/x_{4,j}}
-e^{-i\frac\tau2}\prod_{j=1}^{K_4}\frac{x-x_{4,j}^+}{x-x_{4,j}}\prod_{j=1}^{K_2}\frac{x-x_{2,j}^+}{x-x_{4,j}}\frac{x-1/x_{2,j}^+}{x-1/x_{4,j}}\\
&&=2i\sin(\tau/2)\prod_{j=1}^{K_1}\frac{x-1/x_{1,j}}{x-1/x_{4,j}}\prod_{j=1}^{\t
K_1}\frac{x-1/\tilde x_{1,j}}{x-1/ x_{4,K_1+j}}\prod_{j=1}^{
K_3}\frac{x-x_{3,j}}{x-x_{4,j}}\prod_{j=1}^{\t K_3}\frac{x-\t
x_{3,j}}{x-x_{4,K_3+j}}
\la{eq:fermionic}
\eeqa
In this form it is easy to expand the duality relation in powers of
$1/\sqrt\lambda$. By expanding all factors in (\ref{eq:fermionic}) such
as
$$
\prod_{j=1}^{K_2}\frac{x-x_{2,j}^\pm}{x-x_{4,j}}
=\exp\(\sum_{j=1}^{K_2}\log\frac{x-x_{2,j}^\pm}{x-x_{4,j}}\)\simeq\exp\(\mp\frac{i}{2}G_2(x)+\sum_j^{K_2}\frac{\epsilon_{2,j}}{x-x_{2,j}}\)\;,
$$
we find
\beqa
\nn \sin\(\frac{\eta(p_4-p_3)}{2}\)&=&\sin\(\frac\tau2\)\exp\(+\sum\frac{\epsilon_3}{x-x_3}+\sum\frac{\t\epsilon_3}{x-x_3}-\sum\frac{\epsilon_2}{x-x_2}\)
\\
\nn&\times&\exp\(-\sum\frac{\epsilon_1/x_1^2}{x-1/x_1}-\sum\frac{\t\epsilon_1/\t
x_1^2}{x-1/\t x_1}+\sum\frac{\epsilon_2/x_2^2}{x-1/x_2}\) \,.
\eeqa
Then, similarly to what we had in section \ref{dererivationanomaly} for
the bosonic duality, we notice that
\beqa
\nn \alpha(x)\d_x\(\sum\frac{\epsilon_3}{x-x_3}+\sum\frac{\t\epsilon_3}{x-\t
x_3}-\sum\frac{\epsilon_2}{x-\t x_2}\)=H_3+H_{\t3}-H_4-H_2\;,
\eeqa
with a similar expression for the argument of the second exponential.
Thus finally we get
$$
\(H_4+H_2-H_3-H_{\t 3}\)+\(\bar H_2-\bar H_1-\bar H_{\t
1}\)
=-\cot_{34}\;,
$$
or alternatively, using the $x\to 1/x$ symmetry transformation
properties of the quasi-momenta,
$$
\(\bar H_4+\bar H_2-\bar H_3-\bar H_{\t 3}\)+\(H_2-H_1-H_{\t
1}\)
=-\cot_{12}\;. \la{Hs1}
$$
From this expressions we can deduce several properties of the density
mismatches we wanted to obtain. For example, if we compute the
discontinuity of (\ref{Hs1}) at a cut containing roots $x_1$, that is
in a large cut of stacks $\mathcal{C}_{1,i>4}$, we immediately get
\beqa
\rho_1-\rho_2&=&-\frac{\Delta\cot_{12}}{2\pi i}\;\;,\;\;x\in{\cal
C}_{1,i>4}\la{eq:rho12} \,.
\eeqa
Proceeding in a similar way we find
\beqa
\rho_3-\rho_4&=&-\frac{\Delta\cot_{34}}{2\pi i}\;\;,\;\;x\in{\cal
C}_{3,i>4}\la{eq:rho43}\,,\\
\rho_{3}-\rho_4&=&\rho_2-\rho_{\t 3}\;\;,\;\;x\in{\cal
C}_{1,i>4}\cup{\cal C}_{2,i>4}\la{eq:rho4323}\,.
\eeqa
Let us now show that in the scaling limit the fermionic  duality corresponds
just to the exchange of the sheets $\{p_i\}$ of the Riemann surface.
For illustration let us pick $p_1$ and see how it transforms under the
duality. By definition the fermionic duality corresponds to the
replacement $\eta\to -\eta, H_1\to H_{\t 1}, H_3\to H_{\t 3}$ and
$\phi_1 \leftrightarrow \phi_2$, $\phi_3\leftrightarrow \phi_4$, so
that
\beqa
\nn p_1 \to \displaystyle{\frac{2\pi \J x  -
\delta_{\eta,-1}Q_1+
\delta_{\eta,+1}Q_2 x }{x^2-1}} -\eta\( - H_{\t 1} - \bar H_{\t 3} + \bar
H_4\)+\phi_2=p_2+\eta\cot_{12}
\eeqa
In the same way we get
\beqa
p_2 \to p_1+\eta\cot_{12}\;\;,\;\;
\nn p_3 \to p_4-\eta\cot_{34}\;\;,\;\;p_4 \to p_3-\eta\cot_{34}\;,
\eeqa
and since $\cot_{ij}\sim 1/\sqrt\lambda$ we see that to the leading
order the duality indeed just exchanges the sheets.
\subsubsection{Bosonic duality in scaling limit}
The bosonic nodes of the BS equations are precisely as in the usual
Bethe ansatz discussed in the first sections so that we can just
briefly mention the results. The duality ($\tau=\eta(\phi_2-\phi_3)$)
$$
e^{+i\frac\tau2}\t Q_2(u-i/2)  Q_2(u+i/2)
-e^{-i\frac\tau2}\t Q_2(u+i/2)  Q_2(u-i/2)
=2i\sin\frac{\tau}{2}Q_1(u)Q_3(u)
$$
leads to
\beq
(H_1+H_3-H_2-H_{\t2})+(\bar H_1+\bar H_3-\bar H_2-\bar
H_{\t2})=\cot_{23}\la{eq:bosd}
\eeq
which implies
$$
\rho_2-\rho_3=+\frac{\Delta\cot_{23}}{2\pi i}\;,\;x\in{\cal
C}_{2,i>4}\la{rho23}
$$
\FIGURE[t]{
   \centering
       \resizebox{90mm}{!}{\includegraphics{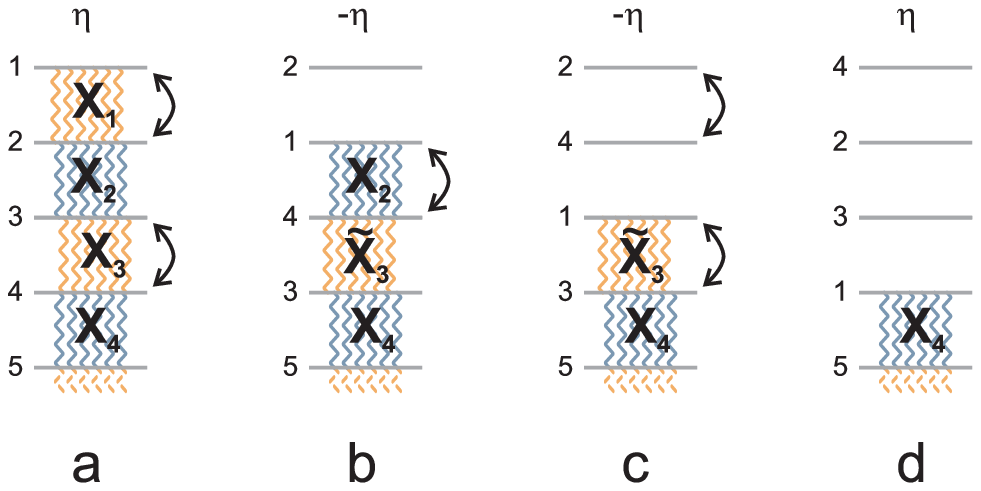}}
       \caption{Action of the duality on a long stack. By
successively applying the fermionic and the bosonic dualities duality
we can reduce the size of any large cut. One should not forget to
change the sign of the grading $\eta$ after applying the fermionic
duality.}
       \la{BFdual}
}

As we already discussed  in section \ref{one} the bosonic duality also amounts to an
exchange of Riemann sheets. Indeed, under the replacement
$H_2\to H_{\t 2}$ and $\phi_2\leftrightarrow \phi_3$, we find
$$
p_2\to p_3-\eta\cot_{23}\;\;,\;\;p_3\to p_2+\eta\cot_{23}
$$
which again, to the leading order in $\sqrt\lambda$, is just the
exchange of the sheets of the curve.
\subsubsection{Dualities and the missing mismatches}

Using bosonic and fermionic dualities separately we already got some
information about the several possible mismatches of the densities
inside the stack. To compute the missing mismatches we have to use both
dualities together. For example suppose we want to compute
$\rho_3-\rho_4$ in a cut $\mathcal{C}_{1,i>4}$. We start by one such
large cut of stacks (see figure \ref{BFdual}a) and we apply the
fermionic duality to this configuration so that we obtain a smaller cut
as depicted in figure \ref{BFdual}b. For this configuration we can use
\eq{rho23} to get
$$
\rho_2-\rho_{\t 3}=+\frac{\Delta\cot_{14}}{2\pi i}\;.
$$
However, from \eq{eq:rho4323}, this is also equal to the mismatch we
wanted to compute, that is
$$
\rho_3-\rho_4=+\frac{\Delta\cot_{14}}{2\pi i}\;\;,\;\;x\in{\cal
C}_{1,i>4}\;.
$$
To compute the last mismatch we apply the bosonic duality to get a yet
smaller cut as in figure \ref{BFdual}c for which we use \eq{eq:rho43}
to get
$$
\rho_{\t 3}-\rho_4=-\frac{\Delta\cot_{13}}{2\pi i}\;.
$$
Again, from \eq{eq:rho4323}, we can revert this result into a mismatch
for the configuration before duality, that is
$$
\rho_{2}-\rho_3=-\frac{\Delta\cot_{13}}{2\pi i}\;\;,\;\;x\in{\cal
C}_{1,i>4}\;.
$$
Let us then summarize all densities mismatches in table \ref{tb:1}.
\TABLE[t]{
\caption{\small\textsf{\textit{Densities missmatches}}}\la{tb:1}
\begin{tabular}{c|l|l|l}
 & ${\cal C}_{1,i}$ & ${\cal C}_{2,i}$ & ${\cal C}_{3,i}$\\\midrule
  $2\pi i(\rho_1-\rho_2)$ & $\;-\Delta\cot_{12}\;$ &\\
  $2\pi i(\rho_2-\rho_3)$ & $\;-\Delta\cot_{13}\;$ & $\;+\Delta\cot_{23}\;$\\
  $2\pi i(\rho_3-\rho_4)$ & $\;+\Delta\cot_{14}\;$ & $\;-\Delta\cot_{24}\;$ & $\;-\Delta\cot_{34}\;$
\end{tabular}
}

\subsection{Integral equation}
In this section we shall recast equation \eq{midan} or
\beq
\eta\frac{4\pi \J  x  - 2\delta_{\eta,+1}{\cal Q}_1 -
2\delta_{\eta,-1}{\cal Q}_2 x }{x^2-1}+2\,\sH_4-H_3-H_5-\bar H_1-\bar
H_7=2\pi n +\eta\phi_4-\eta\phi_5-\cot_{45} \la{Hmid}
\eeq
in terms of the density $\rho_4(x)$ of the middle roots $x_4$. To do so
we only need to
replace the several densities by the middle node density $\rho_4(x)$
using the several density mismatches presented in table \ref{tb:1}.
Defining
$$
H_{ij}(x)\equiv\int\limits_{\mathcal{C}_{ij}}\frac{\alpha(x)}{\alpha(y)}\frac{\rho_4(y)}{x-y}dy
$$
we can then rewrite equation (\ref{Hmid}) in terms of the middle node
roots only,
\beqa
\nn&&\eta\frac{4\pi  \J x  - 2\delta_{\eta,+1}\mathcal{Q}_1-
2\delta_{\eta,-1} \mathcal{Q}_2 x }{x^2-1}+2\
\sH_{45}+H_{15}+H_{48}-2\bar H_{18}-\bar H_{15}-\bar H_{48}
\\ \la{stringeq}
&&=2\pi n+\eta\phi_4-\eta\phi_5-\cot_{45}+
\mathop{\sum_{1\leq i\leq4}}_{5\leq j\leq 8}(\iint_{ij}^{i4}+\iint_{ij}^{5j})
+
\mathop{\sum_{1\leq i\leq4}}_{5\leq j\leq
8}(\bar\iint_{1j}^{i1}+\bar\iint_{i8}^{8j})
\eeqa
where $x\in \mathcal{C}_{45}$ and
$$
\iint_{ij}^{kl}(x)=(-1)^{F_{kl}}\int\limits_{\mathcal{C}_{ij}}\frac{\alpha(x)}{\alpha(y)}\frac{\Delta\cot_{kl}}{x-y}\frac{dy}{2\pi
i}\;\;,\;\;\iint_{ij}^{kk}(x)\equiv
0\;\;,\;\;\bar\iint_{ij}^{kl}(x)=\iint_{ij}^{kl}(1/x) \,.
$$
The several dualities amount to an exchange of Riemann sheets so that the cuts $\mathcal{C}_{ij}\to \mathcal{C}_{i'j'}$ with the subscripts in $H_{ij}$ changing accordingly.
The middle roots $x_4$ are never touched in the process. Moreover to leading order $p_i \leftrightarrow p_{i'}$ and thus the r.h.s. of (\ref{stringeq}) is also trivially changed under the dualities. Therefore, as in section \ref{anomalies} (see (\ref{master}) and (\ref{master13})), we can now trivially write the corrected equation when $x$ belongs to any possible type of cut of stacks by applying the several dualities to equation (\ref{stringeq}).

\subsection{Fluctuations}\la{stringfluctsec}
In this section we shall find the integral equation (\ref{stringeq}) from
the field theoretical point of view like we did in section \ref{1loop}
and in appendix B. That is, we will find what the corrections to the
classical (leading order) equations \cite{BKSZ}
\beq
\eta\frac{4\pi \J x  - 2\delta_{\eta,+1}\mathcal{Q}_1-
2\delta_{\eta,-1}\mathcal{Q}_2 x }{x^2-1}+2\,\sH_4-H_3-H_5-\bar
H_1-\bar H_7=2\pi n+\eta\phi_4-\eta\phi_5\,, \la{leading}
\eeq
\textit{should be in order to describe properly the semi-classical
quantization of the string} (and not only the classical limit). We will
find that this construction leads precisely to the integral equation
\eq{stringeq} thus showing that the BS nested Bethe ansatz equations do
reproduce the $1$-loop shift around any (stable) classical solution
with exponential precision (in some large charge of the classical
solution). This section is very similar to section
\ref{1loope} and to Appendix B and thus we will often omit lengthy but
straightforward intermediate steps. We assume $i=1,\dots,4$ and
$j=5,\dots,8$ in all sums.

As in (\ref{inteqrho}) and (\ref{inteqrhogen}), we add
$\frac{1}{2}(-1)^F$ of a  virtual excitation for each possible mode
number $n$ and polarization $ij$ to each quasi-momenta. Notice that for
this super-symmetric model the fluctuations can also be fermionic and
indeed the grading  $(-1)^{F}$ equals $+1$ ($-1$) for bosonic
(fermionic) fluctuations, see figure \ref{fig:simple}, as usual for
bosonic (fermionic) harmonic oscillators.

We denote $\rho=\rho_0+\delta\rho$ where $\rho_0$ is the leading
density, solution of the leading (classical) equation \eq{leading},
while $\rho$ obeys the corrected (semi-classical) equation. For example, if
we consider $x\in {\cal C}_{4,5}$, the starting point should be (see
\cite{GV2} for a similar analysis)
\beqa
&&\frac{-2x\delta_{\eta,-1}\delta \mathcal{Q}_1
}{x^2-1}+2\int\limits_{{\cal C}_{45}}
\frac{\alpha(x)}{\alpha(y)}\frac{\delta\rho(y)}{x-y}
+
\int\limits_{{\cal C}_{15}} \frac{\alpha(x)}{\alpha(y)}\frac{\delta\rho(y)}{x-y}
+
\int\limits_{{\cal C}_{48}}
\frac{\alpha(x)}{\alpha(y)}\frac{\delta\rho(y)}{x-y}\nn\\
&&-2\int\limits_{{\cal C}_{18}}
\frac{\alpha(1/x)}{\alpha(y)}\frac{\delta\rho(y)}{1/x-y}
-
\int\limits_{{\cal C}_{15}}
\frac{\alpha(1/x)}{\alpha(y)}\frac{\delta\rho(y)}{1/x-y}
-
\int\limits_{{\cal C}_{48}}
\frac{\alpha(1/x)}{\alpha(y)}\frac{\delta\rho(y)}{1/x-y}\nn\\
&&+\sum_{n=-N}^N\frac{1}{2}\[\sum_{i<4}\frac{\alpha(x)}{x-x_n^{i5}}+\sum_{j>5}\frac{\alpha(x)}{x-x_n^{4j}}
-\sum_{i<4}\frac{\alpha(1/x)}{1/x-x_n^{i8}}-\sum_{j>5}\frac{\alpha(1/x)}{1/x-x_n^{1j}}\]=0
\la{inteqdelta}
\eeqa
Then, by construction, the charges
\beq
Q_r=\int_{\cal C}\frac{\rho(y)}{y^r}dy+
\sum_n\sum_{ij}(-1)^{F_{ij}}\frac{\alpha(x_n^{ij})}{2(x_n^{ij})^r}
=\int_{\cal
C}\frac{\rho(y)}{y^r}dy+\sum_{ij}\frac{(-1)^{F_{ij}}}{2}\ccw\oint_{x_n^{ij}}\frac{\cot_{ij}}{y^r}\frac{dy}{2\pi
i} \la{Qrs}
\eeq
will take the $1/\sqrt\lambda$ corrected values. It is clear that, as
before, we do not include the new \textit{virtual} excitations in the
density $\rho(x)$. Similarly to \eq{derdens} and \eq{densAp}, if we
want the charges to have the standard form
$$
\mathcal{Q}_r=\int\frac{\varrho(y)}{y^r}dy
$$
we must redefine the density as
$$
\varrho=\rho+\frac{1}{4\pi i}\(\sum_{i<i'\leq4}(-1)^{F_{ii'}}\Delta\cot_{ii'}
+\sum_{j>j'\geq5}(-1)^{F_{jj'}}\Delta\cot_{jj'}\) \,.
$$
Now we want to go back to the integral equation \eq{inteqdelta} and
rewrite it using the density $\delta\varrho=\varrho-\rho_0$. For
example, for $x\in {\cal C}_{45}$,
\beqa
&&\nn 2\int\limits_{{\cal C}_{45}}
\frac{\alpha(x)}{\alpha(y)}\frac{\delta\rho(y)}{x-y}
+
\int\limits_{{\cal C}_{15}} \frac{\alpha(x)}{\alpha(y)}\frac{\delta\rho(y)}{x-y}
+
\int\limits_{{\cal C}_{48}}
\frac{\alpha(x)}{\alpha(y)}\frac{\delta\rho(y)}{x-y}\\
&&
+\sum_{n=-N}^N\frac{1}{2}\[\sum_{i}\frac{(-1)^{F_{i5}}\alpha(x)}{x-x_n^{i5}}+\sum_{j}\frac{(-1)^{F_{4j}}\alpha(x)}{x-x_n^{4j}}
\]=\nn\\ \nn
&&2\int\limits_{{\cal C}_{45}}
\frac{\alpha(x)}{\alpha(y)}\frac{\delta\varrho(y)}{x-y}
+
\int\limits_{{\cal C}_{15}}
\frac{\alpha(x)}{\alpha(y)}\frac{\delta\varrho(y)}{x-y}
+
\int\limits_{{\cal
C}_{48}}\frac{\alpha(x)}{\alpha(y)}\frac{\delta\varrho(y)}{x-y}\\
&&\nn +
\cot_{45}-\sum_{ij}\(\iint_{ij}^{4i}+\iint_{ij}^{j5}\)
-
\frac{1}{2}\sum_{ij}\(\bar\iint_{8j}^{8 i}+\bar\iint_{1i}^{1
j}+\bar\iint_{ij}^{8 i}+\bar\iint_{ij}^{1 j}\)
\eeqa
where the identity
$$
(-1)^{F_{4i}}\cot_{4,i}=-\sum_j\(\iint_{4j}^{4i}+\iint_{ij}^{4i}\)-\sum_j\(\bar\iint_{1j}^{1\bar
i}+\bar\iint_{\bar ij}^{1\bar i}\) \,,
$$
where $\bar{\bar i}=i,\ \bar 1=4,\ \bar 2= 3$, is being used. Now, when
$x\in {\cal C}_{18}$, we will get
\beqa
&&\nn 2\int\limits_{{\cal C}_{18}}
\frac{\alpha(x)}{\alpha(y)}\frac{\delta\rho(y)}{x-y}
+
\int\limits_{{\cal C}_{15}} \frac{\alpha(x)}{\alpha(y)}\frac{\delta\rho(y)}{x-y}
+
\int\limits_{{\cal C}_{48}}
\frac{\alpha(x)}{\alpha(y)}\frac{\delta\rho(y)}{x-y} \\
&&
+\sum_{n=-N}^N\frac{1}{2}\[\sum_{i}\frac{(-1)^{F_{i8}}\alpha(x)}{x-x_n^{i8}}+\sum_{j}\frac{(-1)^{F_{1j}}\alpha(x)}{x-x_n^{1j}}
\]=\nn\\ \nn
&&2\int\limits_{{\cal C}_{18}}
\frac{\alpha(x)}{\alpha(y)}\frac{\delta\varrho(y)}{x-y}
+
\int\limits_{{\cal C}_{15}}
\frac{\alpha(x)}{\alpha(y)}\frac{\delta\varrho(y)}{x-y}
+
\int\limits_{{\cal
C}_{48}}\frac{\alpha(x)}{\alpha(y)}\frac{\delta\varrho(y)}{x-y} \\
&&\nn-\frac{1}{2}\sum_{ij}\(\iint_{ij}^{1i}+\iint_{ij}^{j8}-\iint_{8j}^{8i}-\iint_{1i}^{1j}\)
\eeqa
Finally we can use the $x$ to $1/x$ symmetry to translate last equality
into one for $x \in \mathcal{C}_{45}$. Subtracting it from the previous
equation we see that the $1/\sqrt{\lambda}$  corrected equation will
correspond to adding
$$
-\cot_{45}+\sum_{ij}(\iint_{ij}^{4i}+\iint_{ij}^{5i}+
\bar\iint_{1j}^{1i}+ \bar\iint_{8i}^{8j})
$$
to the r.h.s. of \eq{leading} thus obtaining, after the identification
$\varrho=\rho_4$, precisely the finite size corrected equation
(\ref{stringeq}) obtained from the NBA point of view!

\subsection{The unit circle and the Hernandez-Lopez phase}
In the last section we showed that the one loop shift as a sum of all
fluctuation energies (or others local charges) perfectly matches the finite size corrections in the NBA equations. However we systematically dropped the
contours around the unit circle.

For example, when we blow the contour in the last term of \eq{Qrs}, we
also get some contribution from the singularities inside the unit
circle. That is we will have an extra contribution to the charges given
by an integral over the unit circle. Also, take (\ref{inteqdelta}) for
instance. To pass to the r.h.s we transformed the collections of poles
into integrals over the excitation points and then we blew the contour
which became a collection of contours on the several existing cuts.
Again we dropped the contribution from the integrals over the unit
circle which would lead to an extra $1/\sqrt{\lambda}$ term in the
r.h.s. of (\ref{stringeq}). In our previous paper \cite{GV2} we
showed\footnote{Recently the HL phase was also found \cite{Chen:2007vs}
in the study of the open string scattering of giant magnons
\cite{Hofman:2006xt}.} that this extra contribution matches precisely
the extra contribution coming from the Hernandez-Lopez phase in the
NBA!

However, as we explained in \cite{GV2}, in
order to obtain precisely the HL phase a precise prescription for the labeling of the mode numbers of the fluctuations must be given.

Moreover, in \cite{GV2}, we assumed that everywhere we can replace $\cot\(\frac{p_i(x)-p_j(x)}{2}\)$ by $i\ {\rm sign}({\rm Im}\ x)$ with exponential precision in $\frac{L}{\sqrt\lambda}$. This is reasonable for generic points in the unit circle, where the imaginary part of $p_i(x)-p_j(x)$ is large, but one has to carefully analyze the neighbourhood of the real axis, where this imaginary part vanishes.

Let us consider these two subtle points in greater detail.

\subsubsection{A mode number prescription}
As we emphasized  in \cite{GV1} if we number the fluctuation charges
$\mathcal{Q}_n^{ij}$ differently we might obtain different results for the $1$-loop shift, that is for the graded sums of these fluctuation charges.
Thus a precise prescription for the labeling of the quantum
fluctuations is crucial. In the appendix A of \cite{GV2} we found out
that the contribution of the integrals of the previous section does
reproduce the HL phase provided we number the quantum fluctuations
located at $x_n^{ij}$ according to
$$
p_i(x_n^{ij})-p_j(x_n^{ij})=2\pi \(n-m_i+m_j\)
$$
with some specific choice of $m_i$. Moreover we also showed that for
the same choice of $m_i$ the contribution to the charges coming from
the above mentioned integrals over the unit circle is zero. Using the
$x$ to $1/x$ symmetries following from the definition of the
quasi-momenta (\ref{eq:p}) plus the restriction (\ref{restriction}) on
the twists, we can redo the computation in the Appendix A of \cite{GV2}
to find that  the condition on the $m_i$ now reads
$$
\(m_2+m_3-m_1-m_4\) \(m_5+m_8-m_6-m_7\)=0
$$
so that, in particular, $m_i=0$ does the job nicely. We see that, with
the introduction of these twists and subsequent redefinition of the
quasimomenta, the prescription for the labeling of the excitations
becomes absolutely natural and algebraic curve friendly
\cite{GV1}. This answers the question raised in \cite{GV2}  concerning the naturalness of the presciption needed to obtain the HL phase \cite{HL}
-- see appendix A in
\cite{GV2}.

\subsubsection{Unit circle contribution}
Let us now us focus on the vicinity of $x=1$ where we have the following
expansion of the quasi-momenta
$$
\frac{p_i(x)-p_j(x)}{2}=\frac{\beta_{ij}}{x-1}+\dots
$$
where $\beta_{ij}$ is usually of order $L/\sqrt\lambda$ (and should be so for the asymptotical BAE to be valid). We will consider the circle
with radius $x_{N+1/2}^{ij}\simeq 1+\frac{1}{\pi N\beta_{ij}}$, where
$N$ is some large cutoff in the sum of fluctuations
\eq{inteqdelta}. We want to estimate
$$
\int\alpha(x)f(x)\[\cot\(\frac{p_i-p_j}{2}\)+i\ {\rm sign}({\rm Im}\ x)\](p'_i-p'_j)dx \,.
$$
This integral is dominated for $x\simeq \pm 1$ and can be performed by saddle point. The contribution for $x\simeq 1$ is
$$
\int\alpha(x)f(x)\[\cot\(\frac{p_i-p_j}{2}\)+i\ {\rm sign}({\rm Im}\ x)\](p'_i-p'_j)dx=
\frac{i\pi^3 f(1)}{6\beta_{ij}\sqrt\lambda}+{\cal O}\(\frac{1}{N}\)
$$
which is zero under the sum over all polarizations. For example
$$
\frac{(-1)^{F_{45}}}{\beta_{45}}=-\frac{(-1)^{F_{35}}}{\beta_{35}}\;.
$$
Thus we can indeed drop the $\cot$'s when integrating over the unit circle and thus we finally conclude that the one loop shift to any local charge computed from the BS equations with the Hernandez-Lopez phase is indeed given by the sum of fluctuations as predicted by field theoretical arguments.

\subsection{Zero twist and large fillings via analytical
continuation}\la{zerotw}

Although we always assumed the twists to be sufficiently large and the
fillings to be sufficiently small we can always analytically continue
the results towards zero twists or large filling fractions. Let us
briefly explain why. In the scaling limit, for large twists, the
bosonic duality we introduced amounts to a simple exchange of sheets in
some Riemann surface, $p_a(x)
\leftrightarrow p_b(x)$. As we saw in section \ref{examples} what
happens when the twists start to become very small is that the
quasi-momenta are still simply exchanged but in a piecewise manner,
that is, we can always split the complex planes in some finite number of regions where the bosonic duality simply means $p_a(x)
\leftrightarrow p_b(x)$.
Thus, from the $e^{ip}$ algebraic curve point of view nothing special
occurs for what analyticity is concerned and therefore we can safely
analytically continue our findings to any value of the twists. Exactly the  same analysis holds for the filling fractions.
Moreover, for the usual Bethe system, we defined a set of
quasi-momenta, which constitute an algebraic curve to any order in
$1/L$, and therefore we don't expect analyticity to break down at any
order in $1/L$.

We also preformed a high precision numerical check
concluding that there is no singularity when the configuration of the
Bethe roots is affected by this partial reshuffling of the sheets and
that finite size corrections are still related to the same sum of
fluctuations, which are analytical functions w.r.t. the twists.

\section{Conclusions}

In this paper we studied generic nested Bethe ansatz (NBA) equations,
the corresponding scaling limit and its leading finite size corrections.
Let us summarize briefly our main results
\begin{itemize}
\item{We found out that the introduction of some extra phases, called twists, are crucial for the formation
of bound states of roots of different types, called in the literature
by stacks \cite{BKSZII}. Strictly speaking without these twists the
stacks do not exist. See sections \ref{one} and \ref{examples}.}
\item{We understood how to use the bosonic duality between various systems of Bethe roots which is present even in the absence of any fermionic symmetry. In the scaling limit we showed that this duality amounts to a reshuffling of Riemann sheets of the algebraic curve formed by the condensation of Bethe roots. See sections \ref{one} and \ref{dualitysec}. }
\item{We explained how to write down the integral equation describing the leading finite size corrections around generic NBA's for (super) spin chains by using the transfer matrices for (super) group along with some $TQ$ relations. See section \ref{Ts}}
\item{We provided an alternative derivation of this integral equation using an independent path, namely using the dualities present in the Bethe equations allowing one to get rid of the several stacks and reduce the size of any cut by successive application of several dualities. See section \ref{dererivationanomaly}.}
\item{We obtained the integral equation describing the finite size corrections to the Beisert-Staudacher equations \cite{BS} with the Hernandez-Lopez phase \cite{FS1,HL} in the scaling limit (to do so we were forced to use the duality approach because at present the $psu(2,2|4)$ transfer matrices for this Bethe ansatz are not known\footnote{See section 6 in \cite{Beisert:2006qh} for some attempts to fill this gap.}). See section \ref{BSsection}.}
\item{In the scaling limit Beisert-Staudacher equations \cite{BS} describe the classical motion of the superstring on $AdS_5\times S^5$ through the finite gap curves of \cite{BKSZ}. Thus the integral equation we found should reproduce the $1$-loop shift for all the charges around any classical string motion and this is obviously a very nontrivial check of the validity of the BS equations. We show that this equation indeed mimics the presence of a sea of virtual particles thus proving this general statement. See section \ref{stringfluctsec}.}
\end{itemize}

\acknowledgments
We would like to thank T.~Bargheer, N.~Beisert, J.~Penedones, A,~Rej,
K.~Sakai, M.~Staudacher, A.~Zabrodin and especially V.~Kazakov for many
useful discussions. The work of N.G. was partially supported by French
Government PhD fellowship, by RSGSS-1124.2003.2 and by RFFI project
grant 06-02-16786. N.G. thanks CFP, where part of this work was done,
for the hospitality during his visit. N.G and P.V thank AEI Potsdam,
where part of this work was done, for the hospitality during the visit.
P.~V. is funded by the Funda\c{c}\~ao para a Ci\^encia e Tecnologia
fellowship {SFRH/BD/17959/2004/0WA9}. P.V. thanks PNPI, where part of
this work was done, for the hospitality during his visit.

\addtocontents{toc}{\protect\contentsline{section}{Appendix A: Transfer matrix invariance and bosonic duality for $SU(K|M)$ supergroups}{\arabic{page}\protect}}
\section*{Appendix A: Transfer matrix invariance and the bosonic duality for $SU(K|M)$ supergroups}\la{invariance}

In this section we review the formalism of \cite{KSZ} which allows one
to derive the transfer matrices of usual (super) spin chains in any
representation. We will use this general formalism to prove the
invariance under the bosonic dualities of all possible transfer
matrices one can build. The transfer matrices presented in section
\ref{Ts} can be obtained trivially using this formalism\footnote{We
should mention that the transfer matrices in section \ref{Ts} are not
exactly the same we have in this Appendix but can be obtained from
these via a trivial rescaling in $u$ which obviously does not spoil the
invariance of these objects.}.

As mentioned in section \ref{one}, for the standard $SU(K|M)$ super
spin chains (based on the standard $R$--matrix $R(u)=u + i \mathcal{P}$
with $\mathcal{P}$ the super permutation) we can find the (twisted)
transfer matrix eigenvalues for the single column young tableau with
$a$ boxes through the \textit{non-commutative generating functions}
\cite{KSZ,Zab}
\beq
\sum_{a=0}^{\infty} (-1)^a e^{i a \partial_u} \frac{T_a(u)}{Q_{K,M}(u+(a-K+M+1)\, i/2 )} e^{i a \partial_u} = \overrightarrow{\prod}_{(x,n)\in \gamma} \hat V^{-1}_{x,n}(u) \la{gen}
\eeq
where $\gamma$ is a path starting from $(M,K)$ and finishing at $(0,0)$
(always approaching this point with each step) in a rectangular lattice of
size $M\times K$ as in figure \ref{figH0}\footnote{Notice that
the path goes in opposite direction compared to the labelling  $a$ of
the Baxter polynomial $Q_a$ used before. In the notation of this section $Q_{k,m}$ corresponds to the node is at position (m,k) in this
lattice.}, $x=(m,k)$ is point in this path and $n=(0,-1)$ or $(-1,0)$
is the unit vector looking along the next step of the path. Each path
describes in this way a possible Dynkin diagram of the $SU(K|M)$ super
group with corners denoting fermionic nodes and straight lines bosonic
ones, see figure \ref{figH0}.  Finally,
\beqa
\nn \hat V^{-1}_{(m,k),(0,-1)}(u)&=&e^{i\phi_k}\, \frac{  Q_{k,m}(u+i(m-k-1)/2)   }{  Q_{k,m}(u+i(m-k+1)/2)   }\frac{     Q_{k-1,m}(u+i(m-k+2)/2)}{    Q_{k-1,m}(u+i(m-k+0)/2)}-e^{ i \partial_u}\\ 
\nn\hat V^{-1}_{(m,k),(-1,0)}(u)&=&\(e^{i\varphi_m}\,\frac{  Q_{k,m-1}(u+i(m-k-2)/2)   }{  Q_{k,m-1}(u+i(m-k+0)/2)    }\frac{   Q_{k,m}(u+i(m-k+1)/2)}{    Q_{k,m}(u+i(m-k-1)/2)}-e^{ i \partial_u}\)^{-1}
\eeqa
where $Q_{k,m}$ is the Baxter polynomial for the roots of the
corresponding node\footnote{$\hat Q_{0,0}$ is normalized to $1$. If we
are considering a spin in the representation where the first Dynkin
node has a nonzero Dynkin label then $Q_{M,K}$ will play the role of
the potential term. In general the situation is more complicated, see
\cite{KSZ}. In any case we are mainly interested in the dualization of
roots which are not momentum carrying thus we need not care about such
matters.} and $\{\phi_k,\varphi_m\}$ are twists introduced in the
transfer matrix \cite{Zab}. Let us then consider a bosonic node like
the one in the middle of figure \ref{figH0} (the \textit{vertical}
bosonic node is treated in the same fashion). If the position of this
node on the $M\times K$ lattice is given by $(m,k)$ then it is obvious
that the only combination containing $Q_{m,k}$ in the right hand side
of (\ref{gen}) comes from the product of $\hat V^{-1}_{(m,k),(-1,0)}(u)
\hat V^{-1}_{(m+1,k),(-1,0)}(u)$ which reads
\beqa
\nn &&\[e^{i\varphi_m+\varphi_{m+1}} \frac{   Q_{k,m+1}(u+i(m-k+2)/2)}{    Q_{k,m+1}(u+i(m-k+0)/2)}\frac{  Q_{k,m-1}(u+i(m-k-2)/2)   }{  Q_{k,m-1}(u+i(m-k+0)/2)    } +e^{ 2 i \partial_u}- \right.\\
&&\nn \left. -\(e^{i \varphi_{m+1}}\frac{  Q_{k,m}(u+i(m-k-1)/2)   }{  Q_{k,m}(u+i(m-k+1)/2)    }\frac{   Q_{k,m+1}(u+i(m-k+2)/2)}{    Q_{k,m+1}(u+i(m-k+0)/2)}+ \right.\right.\\
&&\left.\left.+e^{i \varphi_{m}}\frac{  Q_{k,m-1}(u+i(m-k+0)/2)   }{
Q_{k,m-1}(u+i(m-k+2)/2)    }\frac{   Q_{k,m}(u+i(m-k+3)/2)}{
Q_{k,m}(u+i(m-k+1)/2)}\)e^{ i \partial_u} \]^{-1}
\eeqa
So, if we want to study the bosonic duality on the node ($k,m$) and its
relation with the invariance of several transfer matrices we need
to study the last two lines of this expression. For simplicity let us
shift $u$, omit the subscript $k$ in the Baxter polynomials
$Q_{k,m-1},Q_{k,m},Q_{k,m+1}$ and define the reduced transfer matrix as
\beq
t(u,\varphi_m,\varphi_{m+1})\equiv e^{i \varphi_{m+1}}\frac{ Q_{m}(u-i)
}{  Q_{m}(u)    }\frac{   Q_{m+1}(u+i/2)}{ Q_{m+1}(u-i/2 )}+e^{i
\varphi_{m}}\frac{  Q_{m-1}(u-i/2 )   }{ Q_{m-1}(u+i/2 )    }\frac{
Q_{m}(u+i)}{ Q_{m}(u)} \,. \la{reduced}
\eeq
Notice that the absence of poles at the zeros of $Q_m$ yields precisely
the Bethe equations for this auxiliary node.
\subsection*{Bosonic duality $\Rightarrow$ Transfer matrices invariance}

Thus, to check the invariance of the transfer matrices in all
representations it suffices to verify that the reduced transfer matrix
$t(u,\varphi_m,\varphi_{m+1})$ is invariant under $\varphi_m
\leftrightarrow \varphi_{m+1}$ and $Q_{m} \to \tilde Q_{m}$ where
\beqa
&&\la{dualm} 2i\sin\(\frac{\varphi_{m+1}-\varphi_m}{2}\) Q_{m-1}(u)Q_{m+1}(u)=\\
&&e^{i\frac{\varphi_{m+1}-\varphi_m}{2}}Q_{m}(u-i/2) \tilde
Q_{m}(u+i/2) -e^{-i  \frac{\varphi_{m+1}-\varphi_m}{2}}Q_{m}(u+i/2)
\tilde Q_{m}(u-i/2)  \nn \,.
\eeqa
which can be easily verified. If suffices to replace, in
$t(u,\varphi_m,\varphi_{m+1})$ in (\ref{reduced}),
\begin{eqnarray*}
\frac{Q_{m}(u-i)}{Q_m(u)}&\to& e^{-i(\varphi_{m+1}-\varphi_m)} \frac{\t Q_m(u-i)}{\t Q_m(u)}\\
&&+2 i e^{-i\frac{\varphi_{m+1}-\varphi_m}{2}} \sin\(\frac{\varphi_{m+1}-\varphi_m}{2}\) \frac{Q_{m-1}(u+i/2)Q_{m+1}(u+i/2)}{Q_m(u)\t Q_m(u)} \,, \\
\frac{Q_{m}(u+i)}{Q_m(u)}&\to& e^{+i(\varphi_{m+1}-\varphi_m)} \frac{\t Q_m(u+i)}{\t Q_m(u)}\\
&&-2 i e^{-i\frac{\varphi_{m+1}-\varphi_m}{2}}
\sin\(\frac{\varphi_{m+1}-\varphi_m}{2}\)
\frac{Q_{m-1}(u-i/2)Q_{m+1}(u-i/2)}{Q_m(u)\t Q_m(u)} \,,
\end{eqnarray*}
which are obvious consequences of the bosonic duality.

\subsection*{Transfer matrix invariance $\Rightarrow$ Bosonic duality}
On the other hand suppose we have two solutions of Bethe equations, one
of them characterized by the Baxter polynomials
$\{\dots,Q_{m-1},Q_{m},Q_{m+1},\dots\}$ with twists
$\{\dots,\varphi_m,\varphi_{m+1},\dots$ and another with
$\{\dots,Q_{m-1},\t Q_{m},Q_{m+1},\dots\}$ with twists
$\{\dots,\varphi_{m+1},\varphi_m,\dots\}$ for which the transfer
matrices are the same, that is
\beq
t(u,\varphi_m,\varphi_{m+1})=\t t(u,\varphi_{m+1},\varphi_{m}) \,.
\eeq
Then we can show that these two solutions are related by the bosonic
duality (\ref{dualm}). Indeed if we build the Wronskian\footnote{We would like to thank A.Zabrodin and V.Kazakov for sugesting this nice interpertation for the bosonic duality} like object
\beqa
W(u)\equiv e^{i\frac{\varphi_{m+1}-\varphi_m}{2}}\frac{Q_{m}(u-i/2)
\tilde Q_{m}(u+i/2)}{Q_{m-1}(u)Q_{m+1}(u)} -e^{-i
\frac{\varphi_{m+1}-\varphi_m}{2}}\frac{Q_{m}(u+i/2)
\tilde Q_{m}(u-i/2)}{Q_{m-1}(u)Q_{m+1}(u)}  \nn \,.
\eeqa
we can easily check that
\beqa
&&W(u+i/2)-W(u-i/2) =\nn \\
&&-e^{-i\frac{\varphi_{m+1}+\varphi_m}{2}}\frac{Q_{m}(u) \tilde
Q_{m}(u)}{Q_{m-1}(u-i/2)Q_{m+1}(u+i/2)}\(t(u,\varphi_m,\varphi_{m+1})-\t
t(u,\varphi_{m+1},\varphi_{m})\)=0 \nn
\eeqa
Since by definition $W(u)$ is a rational function this means it must be
a constant. Thus if $\varphi_m\neq \varphi_{m+1}$ we must have $K_m+\t
K_{m}=K_{m}+K_{m+1}$ and the value of $W$ can be read from the large
$u$ behavior. In this way we obtain precisely the bosonic duality
(\ref{dualm}). If $\varphi_m=\varphi_{m+1}$ then we see that $K_m+\t
K_{m}=K_{m}+K_{m+1}+1$ and we will obtain a different value for the
constant $W$ which will correspond to the untwisted bosonic duality
described in section \ref{examplessu2}.

\addtocontents{toc}{\protect\contentsline{section}{Appendix B: Fluctuations for $su(n)$ spin chains}{\arabic{page}\protect}}
\section*{Appendix B: Fluctuations for $su(n)$ spin chains} \la{gene}
In this Appendix we consider a $su(n)$ NBA with the Dynkin labels $V_a$
being $+1$ for a particular $a$ only (the generalization is obvious).
This example is obviously more general than that considered in section
\ref{1loopfluc} and can be a useful warmup for section \ref{stringfluctsec}
where we find the integral equation describing the $AdS_5\times S^5$
$1$--loop quantization. For the spin chain $su(n)$ NBA, in the
classical limit, we will have $n$ quasi-momenta each one above or below
each of the $n-1$ Dynkin nodes\footnote{See figure \ref{fig:simple} for
an example of such pattern for a super group which clearly resembles
$su(8)$.}. We label these quasi-momenta by $p_i$ ($p_j$) with $i,i'$
($j,j'$) taking positive (negative) values for quasi-momenta above
(below) the node for which $V_a\neq 0$. Then let us mention how the
equations in the previous section are generalized.
We consider a \textit{middle node} cut ${\cal C}_{1,-1}$. The analogue
of equation (\ref{inteqrho}) is now
\beq
\la{inteqrhogen}
-\frac{1}{x}+\sum_j\int\limits_{{\cal C}_{1,j}} \frac{\delta\rho(y)}{x-y}
+
\sum_i\int\limits_{{\cal C}_{i,-1}} \frac{\delta\rho(y)}{x-y}
+\sum_{n=-N}^N\frac{1}{2L}\[\sum_{i}\frac{1}{x-x_n^{i,-1}}+\sum_{j}\frac{1}{x-x_n^{1,j}}\]=0
\eeq
and the charges (\ref{Q0}), (\ref{Q1}), (\ref{Q2}) and (\ref{Q3})
become\footnote{as in the previous section, we are ignoring the
regularization of the charges coming from the contribution of the
contour around the origin which would appear in the second line from
opening the contours around the excitation points $x_n^{ij}$.}
\beqa
Q_r-\int\limits_{{\cal C}} \frac{\rho(y)}{y^r}dy&=&
+\sum_n\sum_{ij}\frac{1}{2L}\frac{1}{(x_n^{ij})^r}
=+\frac{1}{2L}\sum_{ij}\frac{1}{2J}\ccw\oint_{x_n^{ij}}\frac{\cot_{ij}}{y^r}\frac{dy}{2\pi i} \\
&=& +\frac{1}{2L}\sum_{ii'j}\cw\oint_{{\cal
C}_{i'j}}\frac{\cot_{ij}}{y^r}\frac{dy}{2\pi i}+
\frac{1}{2L}\sum_{ijj'}\cw\oint_{{\cal C}_{ij'}}\frac{\cot_{ij}}{y^r}\frac{dy}{2\pi i}\\
&=& -\frac{1}{2L}\sum_{ii'j}\cw\oint_{{\cal
C}_{i'j}}\frac{\cot_{ii'}}{y^r}\frac{dy}{2\pi i}-
\frac{1}{2L}\sum_{ijj'}\cw\oint_{{\cal C}_{ij'}}\frac{\cot_{jj'}}{y^r}\frac{dy}{2\pi i}\\
&=& -\frac{1}{2L}\int_{{\cal
C}}\frac{\sum_{i<i'}\Delta\cot_{ii'}+\sum_{j<j'}\Delta\cot_{jj'}}{y^r}\frac{dy}{2\pi
i} \,,
\eeqa
so that the natural definition of the dressed density becomes now
\beq\la{densAp}
\varrho=\rho+\frac{1}{4L\pi i}\Delta\(\sum_{i<i'}\cot_{ii'}+\sum_{j<j'}\cot_{jj'}\) \,.
\eeq
Next step is to rewrite the integral equation (\ref{inteqrhogen}) in
terms of this new density. We proceed exactly as in (\ref{23}),
(\ref{13}) using now
$$
\cot_{1,i}=-\sum_{j}\(\iint_{{1,j}}^{1i}+\iint_{{i,j}}^{1i}\)\,\,\, , \,\,\, {\iint}_{ij}^{kl}\equiv \int_{{\cal C}_{ij}}\frac{\cot_{kl}(y)}{x-y}\frac{dy}{2\pi i}\,,
$$
which is the analog of (\ref{cot12}) for this $su(n)$ setup, so that at
the end we obtain the following equation
\beq
\sum_j\int\limits_{{\cal C}_{1,j}} \frac{\delta\varrho(y)}{x-y}
+
\sum_i\int\limits_{{\cal C}_{i,-1}} \frac{\delta\varrho(y)}{x-y}
+\frac{1}{L}\(\cot_{1,-1}-\sum_{ij}\int_{{\cal C}_{ij}}\frac{\Delta\cot_{1,i}+\Delta\cot_{j,-1}}{x-y}\frac{dy}{2\pi i}\)=0
\eeq
 for $\delta \varrho=\varrho-\varrho_0$ where $\varrho_0$ obeys the leading order equation
\beq
-\frac{1}{x}+\sum_j\int\limits_{{\cal C}_{1,j}} \frac{ \varrho_0(y)}{x-y}
+
\sum_i\int\limits_{{\cal C}_{i,-1}} \frac{\varrho_0(y)}{x-y}
=2\pi k_{1,-1}\,.
\eeq
This corrected equation is precisely the one we would obtain from
finite size corrections to the $su(n)$ NBA equations. To find this
equation from the Bethe ansatz point of view one can simply repeat
either of the derivations in section \ref{anomalies}, that is the known transfer matrices in various representations or
the bosonic duality described in the previous sections. In section
\ref{BSsection} we consider the AdS/CFT Bethe ansatz equations which
are based on a large rank symmetry group, namely $PSU(2,2|4)$. There
one can see an example of how this could be done in practice (we will
only use the dualities approach because at present we don't have the
$PSU(2,2|4)$ transfer matrices for this (exotic) Bethe ansatz
equations.).


\end{document}